\documentclass[10pt,journal,compsoc]{IEEEtran}
\ifCLASSOPTIONcompsoc
  \usepackage[nocompress]{cite}
\else
  \usepackage{cite}
\fi
\usepackage{url}
\usepackage{textcomp}

\ifCLASSINFOpdf
  \usepackage[pdftex]{graphicx}
\else
\fi
\usepackage{array, caption, tabularx, ragged2e,  booktabs}
\usepackage{xtab}
\usepackage{todonotes}
\usepackage{enumitem}
\usepackage{fixltx2e}

\usepackage{stfloats}
\usepackage{url}

\begin{document}
%
\title{The human side of Software Engineering Teams: an investigation of contemporary challenges} 
%
%
%
%

\author{Marco Hoffmann,
        Daniel Mendez, 
        Fabian Fagerholm
        and Anton Luckhardt
\IEEEcompsocitemizethanks{\IEEEcompsocthanksitem M. Hoffmann was with QualityMinds GmbH, Germany, at the time of writing.\protect\\
E-mail: marco@acureus.com
\IEEEcompsocthanksitem D. Mendez is with the Blekinge Institute of Technology, Sweden and fortiss GmbH, Germany. F. Fagerholm is with Aalto University, Finland and the Blekinge Institute of Technology, Sweden. A. Luckhardt is affiliated with the Technical University of Munich and fortiss GmbH, Germany.}
\thanks{
}}

\IEEEtitleabstractindextext{%
\begin{abstract}

\textbf{Context:} There have been numerous recent calls for research on the human side of software engineering and its impact on various factors such as productivity, developer happiness and project success. An analysis of which challenges in software engineering teams are most frequent is still missing. As teams are more international, it is more frequent that their members have different human values as well as different communication habits. Additionally, virtual team setups (working geographically separated, remote communication using digital tools and frequently changing team members) are increasingly prevalent.

\textbf{Objective:} We aim to provide a starting point for a theory about contemporary human challenges in teams and their causes in software engineering. To do so, we look to establish a reusable set of challenges and start out by investigating the effect of team virtualization. Virtual teams often use digital communication and consist of members with different nationalities that may have more divergent human values due to cultural differences compared to single nationality teams.

\textbf{Method:} We designed a survey instrument and asked respondents to assess the frequency and criticality of a set of challenges, separated in context "within teams" as well as "between teams and clients", compiled from previous empirical work, blog posts, and pilot survey feedback. For the team challenges, we asked if mitigation measures were already in place to tackle the challenge. Respondents were also asked to provide information about their team setup. The survey included the Personal Value Questionnaire to measure Schwartz human values. Finally, respondents were asked if there were additional challenges at their workplace. The survey was first piloted and then distributed to professionals working in software engineering teams via social networking sites and personal business networks.

\textbf{Result:} In this article, we report on the results obtained from 192 respondents. We present a set of challenges that takes the survey feedback into account and introduce two categories of challenges; "interpersonal" and "intrapersonal". We found no evidence for links between human values and challenges. We found some significant links between the number of distinct nationalities in a team and certain challenges, with less frequent and critical challenges occurring if 2-3 different nationalities were present compared to a team having members of just one nationality or more than three. A higher degree of virtualization seems to increase the frequency of some human challenges, which warrants further research about how to improve working processes when teams work from remote or in a distributed fashion.

\textbf{Conclusion:} We present a set of human challenges in software engineering that can be used for further research on causes and mitigation measures, which serves as our starting point for a theory about causes of contemporary human challenges in software engineering teams. We report on evidence that a higher degree of virtualization of teams leads to an increase of certain challenges. This warrants further research to gather more evidence and test countermeasures, such as whether the employment of virtual reality software incorporating facial expressions and movements can help establish a less detached way of communication.
\end{abstract}

}


\IEEEpubid{{\begin{minipage}{\textwidth}\ \\[12pt] \centering
  \copyright 2022 IEEE.  Personal use of this material is permitted.  Permission from IEEE must be obtained for all other uses, in any current or future media, including reprinting/republishing this material for advertising or promotional purposes, creating new collective works, for resale or redistribution to servers or lists, or reuse of any copyrighted component of this work in other works.
\end{minipage}}}

\maketitle

\IEEEdisplaynontitleabstractindextext


%
\IEEEpeerreviewmaketitle

\IEEEraisesectionheading{\section{Introduction}\label{sec:introduction}}
\IEEEPARstart{S}{oftware engineering} is a complex activity with many challenges: Tasks are often novel and require to interpret and understand concepts from various fields, possibly presented by non-technical people. Solutions need to be creatively adapted to different stakeholder requirements. Workers from interdisciplinary fields have to come together to build a product, synchronizing their different points of view in the process. In parallel to this cognitively demanding task, the involved need to exercise human social skills: They need to communicate and understand each other. They should feel safe, appreciated and supported and not be additionally constrained by social challenges; not only in the interest of their own well-being, but also because a happy developer is better at solving problems \cite{graziotin2014happy}.

There have been numerous calls for more research on the human side of software engineering \cite{capretz2014bringing, Whittle2019, fazli2017cultural}. As of now, multiple and diverse human challenge areas have been identified: Improving communication \cite{Lenberg2015}, aiming for more diversity in teams \cite{Menezes2018, berg2012anonymity}, fears or confidence issues \cite{Conboy2011}, team culture and how team members treat each other \cite{wagner2018systematic}, to name a few. The findings so far provide great insights about which areas could matter most and how social challenges in software engineering should be tackled. However, what is still missing is a comprehensive list of most frequent challenges. These challenges should be expressed in a level of detail that is not too deep, so that it does not apply to specific cases only, but also not too abstract, so that it allows pin-pointing of possible improvements. Given such a list of frequent challenges and possible causes -- Why do these challenges occur in that environment? -- could be explored. We assume that understanding the causes will greatly help in finding mitigations. It is our vision to establish a solid list of most relevant human-focused triples (causes, challenge, mitigations) that is continually strengthened, extended, and refined.

In this article, we report on the first step towards this vision. Using survey research, we asked 192 professionals about the frequency, criticality, and mitigation strategies regarding our initial catalogue of human challenges, contextualized twofold: in the context of challenges within software engineering teams, which is the primary focus of this report, but also in the context of challenges occurring between teams and clients of the team. We compiled the list from challenges recorded by the Naming the Pain in Requirements Engineering initiative (short: NaPiRE), which constitutes a comprehensive requirements engineering study on practices and challenges experienced in practical environments \cite{mendez2016naming}~\footnote{See also \url{www.napire.org}.}, as well as from grey literature, such as blog posts. We examined possible influences on the occurrence and perceived criticality of these challenges. We focused on aspects related to global, multinational teams: They often work in virtual teams and they often have many nationalities \cite{han2016framing}. A virtual team refers to teams whose workers are geographically distributed and use digital means to communicate \cite{de2015open}. Geographical distance often implies temporal distance, and thus phases of unavailability of certain team members. This leads to three categories of possible influences: Degree of team virtualization, nation count (as a proxy for potential cultural clashes) and human values.
Furthermore, we used the feedback to refine the initial list and categorize it. This refined list can serve as a future basis for research. Finally, the survey respondents reported on mitigation strategies already in place. This could, in the current state, serve as an inspiration for fellow software engineers.

\subsection{Contribution}
Summarising, these are our contributions and key findings:
\begin{enumerate}
\item A set of relevant human challenges in software engineering teams, as well as a refined set of challenges introducing two categories: challenges that arise between colleagues (interpersonal), and challenges that exist between an individual and their work only (intrapersonal).
\item Insights into how different virtualization aspects of teams influence these challenges - a stronger degree of virtualization may lead to more frequent and critical challenges.
\item Evidence for a possible link between the degree of multi-nationality of a team and our set of challenges. The sweet spot seems to be between 2-5 nationalities.
\item Indicators that correlations between fine-grained human values as defined by Schwartz \cite{schwartz2012refining} and challenge occurrence and criticality in software development are weak.
\item A list of aggregated reported mitigation measures already in place in professional environments for each team challenge (see Appendix \ref{appendix_miti}).
\end{enumerate}

To increase the transparency and reproducibility of our study, we further disclose the whole response dataset, the list of  modifications  done  to create  our  refined  challenges,  as  well  as  descriptions  for  the refined challenges as part of an open data and material set (see Appendix \ref{appendix_repo}).

Using the \emph{Who, What, How sociotechnical framework} by Storey at al. \cite{storey2020software}, we position this work as follows:

\begin{itemize}
    \item \textbf{Who} benefits from this work?
    \begin{itemize}
        \item \emph{Researchers}. The revised problem catalogue plus problem descriptions can be used in further surveys. We report on some interesting links that could be interesting for further research.
        \item \emph{Human}. We collected a list of mitigation measures in place for this first set of relevant human challenges, which can be used by professionals as a reference for potential improvements in their own environment.
    \end{itemize}
    \item \textbf{What} is the main research contribution type?
    \begin{itemize}
        \item \emph{Descriptive}. This article is a first step in finding out which human challenges are most frequent in software engineering, and which social and human aspects have an impact on them.
    \end{itemize}
    \item \textbf{How} is the research conducted?
    \begin{itemize}
        \item \emph{Respondent}. We used survey research to gather our data.
    \end{itemize}
\end{itemize}

\subsection{Outline}
The article is organised as follows. In Section \ref{sec_related}, we describe related work. Section \ref{sec_design} introduces the study design, elaborates on the research questions and the data collection as well as the analysis procedures. In Section \ref{sec_results}, we report on our study results, before concluding our article in Section \ref{sec_conclusion}.
\section{Related Work}\label{sec_related}

Challenges in software engineering teams are widely researched. Culturally induced problems were the most frequent problem type reported in interviews with software professionals \cite{beecham2013we}. Cultural distance may lead to misunderstanding and misinterpretations about things that were said and done in a project, as reported by many case studies investigating Chinese professionals \cite{yugendhar2017comprehensive, foster2018students, wang2017cultural, huang2007cultural}, Irish teams outsourcing parts of their workforce to other countries \cite{casey2004practical, casey2009leveraging}  and other teams of various culture backgrounds \cite{borchers2003software, alsanoosy2018cultural}. Misunderstandings and conversational dominance of specific interlocutors due to varying levels of language proficiency can however be managed to a degree if they are recognised, e.g. by using humour or giving less dominant team members dedicated speaking slots \cite{tenzer2015leading, tenzer2017influence, neeley2015global}.
Some authors advise to keep human value distance low to avoid problems \cite{paasivaara2010practical, liang2007effect}. Others advocate to overcome potential problems in a more productive way: multicultural and diverse teams can perform especially outstandingly \cite{berg2012anonymity, Menezes2018} e.g. if the team tries to actively understand the differences between its members \cite{han2016framing, henderson2018cultural} - a prospect we  want to support and enable.

Globally distributed teams face some additional challenges: If teams in different places are sized differently, there might be a disbalance in perceived power potentially leading to dissatisfaction in the smaller team, or to resentment of the smaller team in the larger team \cite{neeley2015global}. It is harder to develop a relationship of trust and maintain a certain communication frequency \cite{de2015open, neeley2015global}. Communication is often poorer as important cues such as facial expressions are taken away or lack detail when digital means are used \cite{de2015open}. If the language proficiency varies, preferences for used media might differ (with proficient speakers resorting to verbal communication, while less proficient speakers stick to text-based communication) and result in the emergence of isolated social groups \cite{klitmoller2015speaking}.

The impact of certain factors on various performance metrics of teams has some solid groundwork. Almomani et al. conducted an empirical study examining the link between human factors and software process improvement (SPI) in small and medium malaysian teams \cite{almomani2018empirical}. They identified eleven human factors and asked professionals to rate each factor's relevance regarding SPI. The three most relevant factors were leadership involvement, employee involvement and support from senior management. However, the eleven factors seem too coarse and too open to interpretation - support from management can refer to technical, emotional or financial support. Maybe it refers to accessible knowledge - could it be solved with more documentation?  Our challenge set is more fine-grained; thus we expect it to be easier to employ countermeasures if a challenge of the set has been identified.
In a recent systematic literature review, Wagner and Ruhe aggregated evidence about factors that impact the productivity in software development \cite{wagner2018systematic}. They found many human factors that have an impact on productivity, grouped into five categories: "Corporate culture", "Team culture", "Capabilities and experience", "Environment" and "Project". While the existence of evidence for a link between human factors and software engineering productivity and successful software projects is reassuring, the authors do not comprehensively conclude which factors are most relevant and what could be done to improve the chances of success. Although the review itself is from 2018, the most recent paper that was reviewed seems to be from 2008 --- and more than half are older than 2000. We believe that we can contribute with a problem analysis that is up to date for a field that moves very quickly.

In a recent study, Storey et al. investigated the link between developer satisfaction and various factors, with perceived productivity turning out to be one of the most relevant in their data \cite{storey2019towards}. They identified 30 factors that may have an impact on developer satisfaction, and an additional 15 technical and human challenges. For the satisfaction factors, they asked about the importance of each factor, as well as how satisfied they were with each factor. A good manager, high perceived productivity and rewards were considered the three most important factors. For the challenges, respondents had to rate the impact of the presence of each challenge, with poor software architecture, legacy code and the task of finding relevant information being most impactful. 

Spichkova et al. propose some interesting ideas to engineer human factors into software reliability engineering \cite{spichkova2015human}. They suggest to examine human factors such as persons' backgrounds or similarities between persons into test prioritisation, as similar persons potentially make similar mistakes. For requirements engineering, they argue that the detection of inconsistency types may depend on a person's learning style. Having diversity in the group of people validating requirements may lead to better inconsistency detection.

The gap we have identified is a lack of a holistic theory about which attributes of a software engineering team cause which kind of human challenges that hamper a team's ability to perform well, and how the overcoming of these challenges can be supported. Virtual Team settings and difference in personal values are often identified as triggers for challenges, however, it is not yet clear what kind of challenges are affected. In order to build that theory, we believe it is necessary to first identify a set of challenges that have the right level of detail: If a challenge is described too vaguely, it might be affected by many causes, and mitigation will be difficult. If it is too fine-grained, the challenge may only apply to very specific team setups and contexts, and it loses generality and may be subject to changes in working habits too quickly. Using this set of challenges, causes need to be examined and mitigation measures tested. We intend to take a step towards closing this gap by contributing this set of human challenges along with an overview of mitigation measures already in place in the field. Additionally, we provide evidence about the impact of a first set of causes on the occurrence of these challenges. Future studies and surveys can build up on that to find out which aspects in a team matter most.

\subsection{Human Values}\label{rel_persval}
One of our research objectives was to examine the link between human values and the perception of human challenges in the team. While different human value models put emphasis on different aspects in their attempt to measure human values, they all return a quantification of what is important to humans, expressed by multiple dimensions, contextualized by cultural influences. If a link between these values and SE challenges could be found, this would be an important step towards identifying good practises for working together in culturally diverse teams, whose individuals may react very differently to certain situations. Examples of available models for human values are the GLOBE culture dimensions \cite{house2004culture}, the Lewis Model \cite{lewis1999cross} and the Hofstede Model \cite{hofstede2005cultures}. We used the Schwartz human values \cite{schwartz2012refining} measured by the 57-item Portrait Value Questionnaire (PVQ-57RR), which has been validated with representative samples from several countries \cite{schwartz2012refining}. The model was suggested in a recent call to consider human values in software engineering \cite{Whittle2019}. We hypothesize that globally cooperating teams are culturally more diverse – with teams in multiple countries – and will likely work as virtual teams and remote, since high-frequent travel between locations is not feasible. The links between these two traits and challenge occurrence and perception are thus of particular interest to us. However, models such as the GLOBE dimensions or the Hofstede Model are only valid at a national level and were not designed for use at an individual level \cite{brewer2012misuse,mcsweeney2013fashion}.
The PVQ-57RR asks respondents to indicate how similar a person description is to themselves, measuring 19 different human value dimensions:

\begin{itemize}
\item \textbf{Self-direction Thought}—Freedom to cultivate one's own ideas and abilities.
\item \textbf{Self-direction Action}—Freedom  to determine one own's actions.
\item \textbf{Tradition}—Maintaining and preserving cultural, family or religious traditions.
\item \textbf{Stimulation}—Excitement, novelty and change.
\item \textbf{Hedonism}—Pleasure and sensuous gratification.
\item \textbf{Achievement}—Success according to social standards.
\item \textbf{Power Dominance}—Power through exercising control over people.
\item \textbf{Power Resources}—Power through control of material and social resources.
\item \textbf{Face}—Maintaining one's public image and avoiding humiliation.
\item \textbf{Security Personal}—Safety in one's immediate environment.
\item \textbf{Security Societal}—Safety and stability in the wider society.
\item \textbf{Conformity-Rules}—Compliance with rules, laws and formal obligations.
\item \textbf{Conformity-Interpersonal}—Avoidance of upsetting or harming other people.
\item \textbf{Humility}—Recognizing one's insignificance in the larger scheme of things.
\item \textbf{Universalism-Nature}—Preservation of the natural environment.
\item \textbf{Universalism-Concern}—Commitment to equality, justice and protection for all people.
\item \textbf{Universalism-Tolerance}—Acceptance and understanding of those who are different from oneself.
\item \textbf{Benevolence-Care}—Devotion to the welfare of the own social circle or group.
\item \textbf{Benevolence-Dependability}—Being a reliable and trustworthy member of a group or society.
\end{itemize}
\section{Study Design}\label{sec_design}
We used online survey research to reach a wide variety of people working in software engineering teams. In the following, we introduce the study design relevant to the analysis present in this article.
\subsection{Research Questions}
Our objective is to build a catalogue of human challenges with a reasonable level of detail. In order to rate the impact of each challenge, we asked the respondents about the frequency and criticality (i.e. how bad the challenge is) of each challenge in their team. To this end, we formulated research questions to guide the design of the study:

\begin{itemize}
    \item \textbf{RQ1.} What human challenges occur in SE teams?
    \begin{itemize}
        \item \textbf{RQ1.1.} How frequent are these challenges?
        \item \textbf{RQ1.2.} How critical are these challenges?
    \end{itemize}
    \item \textbf{RQ2.} Do individuals’ values influence frequency and criticality of challenges within a team?
    \item \textbf{RQ3.} Does national diversity influence frequency and criticality of challenges within a team?
    \item \textbf{RQ4.} Does the degree of team virtualization influence frequency and criticality of challenges within a team?
\end{itemize}

\subsection{Instrument}
The overall instrument used constitutes in total 260 questions as a maximum used to collect data about the demographics of the respondent [RQ3], characteristics of their team [RQ4], and their human values as measured by the Schwartz Personal Value Questionnaire (PVQ) [RQ2] \cite{schwartz2012refining}. Then, the respondent was asked to report on the frequency and criticality of challenges that can happen in a team and between team and clients [RQ1]. Additionally, we asked if there were mitigation measures in place for the team challenges and if yes, which ones. Finally, at the end of the survey, the respondent could provide feedback about the listed challenges as well as general feedback about the survey and leave their email address if they wanted to receive a notification of the survey results.

\subsubsection{Demographics}
The demographic section contained 14 questions. In Q1, we asked about which gender pronouns the respondent wanted to be addressed with because the PVQ used in the human values section (refer to \ref{subsub_persval}) is adapted to different personal pronouns. Q2-Q6 are questions aimed at finding out about the national diversity of the team: Most important to us was the nationality of the respondent. Q3 was asked because if they had lived in a country for a very long time already, they had probably adapted to the local customs and habits and the original nationality could maybe not be as important. We were also interested in team sizes because in smaller teams, personal differences might have a stronger effect as the teams are more cohesive, with members having more interactions, being more committed to their team and more acquainted with their team members \cite{Bradner2003, Aube2011}. 

Six questions in Q7 ask about the degree of virtualization of the team. We differentiated between "offshore" and "remote" - we define remote work as work that is done at home or any other place that is not a facility of the employer, while offshore refers to teams that work in different countries. Virtualization of a team can be expressed in multiple ways, with likely different implications for each dimension and not all of them necessarily holding true for a virtual team. Each question could either be answered with "Yes" or "No" (see Table \ref{demotable}).

Finally, Q9 determined whether questions about client challenges were displayed to respondents (refer to section \ref{subsub_clientprob}). 

\captionof{table}{Demographic Questions. SC = Single Choice.}
\begin{center}
\label{demotable}
\begin{xtabular}{lp{.335\textwidth}l}
  \midrule
  No. & Question & Type  \\ [0.5ex]
  \midrule
  Q1  & How do you want to be addressed? & SC  \\
  \addlinespace[0.3ex]
  Q2  & In which country are you working? & Open  \\
  \addlinespace[0.3ex]
  Q3  & For how many years have you been working in SE in this country?   & Open  \\
  \addlinespace[0.3ex]
  Q4  & What are your nationalities?   & Open  \\
  \addlinespace[0.3ex]
  Q5  & How many people do you work with?   & Open  \\
  \addlinespace[0.3ex]
  Q6  & What are the nationalities of your team colleagues?   & Open  \\
  \midrule
  Q7.1  & Are you part of an offshore team?   & SC  \\
  \addlinespace[0.3ex]
  Q7.2  & Do you work with offshore teams?   & SC  \\
  \addlinespace[0.3ex]
  Q7.3  & Do you regularly work from remote?   & SC  \\
  \addlinespace[0.3ex]
  Q7.4  & Are your team members geographically dispersed?   & SC  \\
  \addlinespace[0.3ex]
  Q7.5  & Does your team regularly collaborate via communication technology?   & SC  \\
  \addlinespace[0.3ex]
  Q7.6  & Is your team occasionally recomposed?   & SC  \\
  \midrule
  Q8  & How would you describe your role?  & Open  \\
  \addlinespace[0.3ex]
  Q9  & Do you have direct interaction with clients?  & SC  \\
\end{xtabular}
\end{center}

\subsubsection{Human Values}\label{subsub_persval}
To measure human values values as defined by Schwartz \cite{schwartz2012refining}, we used the Portrait Value Questionnaire with 57 items (PVQ-57RR) \cite{schwartz2012refining} (see also section \ref{rel_persval}). The instrument was available with both male and female pronouns and was displayed according to the pronoun preference asked in the demographics section. The instrument asks multiple questions per human value dimension in different wordings and also provides a method to check whether the responses per dimension are coherent. If they are incoherent, the responses can be marked as implausible and be discarded.

\subsubsection{Team Challenges}\label{subsub_teamprob}
We aim to iteratively create a list of human challenges in software engineering by incorporating feedback collected in this and subsequent surveys. To create a starting point for this survey to build upon, we compiled a list of 33 human challenges with a focus on challenges happening "within a company", i.e. problems occurring in teams or in individuals in the context of a team (see Table \ref{teamProbShort}). We used challenges that came up in research about requirements engineering: T1-T20 and T29-T30 were taken from the NaPiRE data \cite{mendez2016naming}. NaPiRE (Naming the Pain in Requirements Engineering) constitutes an initiative by the empirical requirements engineering research community that operates a globally distributed family of survey replications maintained to understand the status quo in practices as well as challenges and their causes and effects in requirements engineering. Details (including publications and open data sets) on the initiative can be taken from the website \url{www.napire.org}. Even more so with agile approaches, requirements engineering is an intertwined part of software engineering and, thus, many of the reported challenges also apply to other phases in software engineering. The unique features of software engineering work — breaking down required solutions into manageable small problems, reaching a shared understanding of complex problems in the team, and constant necessary synchronization about the project and task progress and feasibility  — are shared by requirements engineering. We thus believe this is a reasonable initial list for software engineering challenges. These challenges were then filtered for human-caused challenges as part of a bachelor thesis \cite{Luckhardt2019} and extended by reading grey literature, primarily blog posts, about typical challenges in software engineering teams. T21-T28 were compiled from these \cite{Flint2016, Renee2019, Hammond2016, Batista2019}. T31-T33 were added as result from feedback gathered during the pilot run of the survey. The challenges were then reviewed by the fourth author of this article (who was the main author of \cite{Luckhardt2019}) whether they can be considered a "human" challenge.

For this initial list, we aimed to minimize bias and thus included challenges that may seem more technical than "human", such as "Missing documentation of the project". These are, however, still tasks carried out by humans. Also, that way, comparability with the results of the NaPiRE data is enabled.

To answer RQ1, for each of those challenges, we asked about:
\begin{itemize}
    \item The frequency of it occurring on a 5-point Likert scale, ranging from "Never" to "Very often"
    \item If the challenge happened more often than "Never", we asked about the criticality of the challenge; the danger of it affecting the project's success. For this we employed a 4-point Likert scale ranging from "Not problematic" to "Highly problematic".
    \item Whether there were mitigation measures in place to prevent this challenge. This was a simple optional Yes or No question. 
    \item If there was a measure in place, respondents were asked to describe it using free text.
\end{itemize}

Additionally, respondents were informed that the context of the questions was "challenges within software engineering teams" at the top of the page. In total, a maximum of 132 questions was asked in this section.

\captionof{table}{Team Challenges}
\small
\begin{center}
\begin{xtabular}{lp{.39\textwidth}}
  \midrule
  No. & Team Challenge  \\ [0.5ex]
  \midrule
T1  & Lack of experience\\
T2  & Lack of qualification\\
T3
  & Lack of teamwork skills\\
T4  & Lack of leadership\\
T5  & Communication plan is neglected\\
T6  & Business needs are neglected\\
T7  & Demotivation\\
T8  & Gold plating\\
T9  & Missing willingness to adapt to new software\\
T10  & Missing willingness to adapt to new infrastructure\\
T11  & Missing willingness to adapt to new processes\\
T12  & Insufficient collaboration\\
T13  & Insufficient analysis at the beginning of a task\\
T14  & Subjective interpretations of tasks\\
T15  & Work is not solution-oriented\\
T16  & Language barriers\\
T17  & Conflicts of interests at management level\\
T18  & Missing documentation of the project\\
T19  & Information is not made known to the team\\
T20  & Misunderstandings in communication\\
T21  & Certain people dominating discussions\\
T22  & Missing respect in the workplace\\
T23  & Missing acceptance of alternative lifestyles\\
T24  & Harassment\\
T25  & Slow decision making\\
T26  & Conflicting working styles\\
T27  & People getting angry in discussions\\
T28  & People crying in discussions\\
T29  & Frequently changing team members\\
T30  & People not reporting problems in time\\
T31  & Over-Confidence\\
T32  & Pressure from client forwarded to SE team\\
T33  & Exaggerated seeking of project problems\\
\end{xtabular}
\label{teamProbShort}
\end{center}

\subsubsection{Client Challenges}\label{subsub_clientprob}
We additionally compiled a list of 27 challenges that happen in a context between representatives of the company and external entities, i.e. clients (see Table \ref{clientTableShort}).  We used the same procedures as for the team problems. C1-C21 and C24 were taken from the NaPiRE data \cite{mendez2016naming}, C22 was added as a reverse situation to C21 and the remaining challenges were added as result from feedback gathered during the pilot run of the survey. The challenges were only presented to the respondent if they reported they were working with clients before. For each of these challenges, we asked about the frequency and the criticality as described in \ref{subsub_teamprob}. We did not ask for mitigation measures for the client challenges as to not extend the survey any longer than it already was.
 
Additionally, respondents were informed that the context of the questions was "challenges between teams and clients" at the top of the page. Overall, a maximum of 54 questions was displayed.

\captionof{table}{Client Challenges}
\small
\begin{center}
\begin{xtabular}{lp{.4\textwidth}}
  \midrule
  No. & Client Challenge  \\ [0.5ex]
  \midrule
  C1 & Lack of communication \\
  C2  & Client does not know what they want\\
C3  & Lack of interest in the project by the client\\
C4  & Missing IT project experience at client side\\
C5  & Missing technical knowledge at client side\\
C6  & Exaggerated quality expectation of the client\\
C7  & Conflicts of interests at client side\\
C8  & Client unable to specify functional requirements\\
C9  & Client unable to specify non-functional reqs\\
C10  & Unclear roles and responsibilities at client side\\
C11  & Lack of prioritization by client\\
C12  & Weak management at client side\\
C13  & Insufficient collaboration\\
C14  & Insufficient analysis at the beginning of the project\\
C15  & No direct communication with client\\
C16  & Communication plan is neglected\\
C17  & Subjective interpretations of tasks\\
C18  & Work is not solution-oriented\\
C19  & Language barriers\\
C20  & Misinterpretations\\
C21  & Missing respect towards the client from the team\\
C22  & Missing respect towards the team from the client\\
C23  & Delays due to dependencies to client's third parties\\
C24  & People not reporting problems in time\\
C25  & Over-Confidence\\
C26  & Communication of problems to client restricted\\
C27  & Exaggerated seeking of project problems\\
\end{xtabular}
\label{clientTableShort}
\end{center}

\subsection{Data Collection}
Before the execution of the main survey, we ran a pilot survey with affiliated researchers and software engineers from Germany and Sweden. 11 out of 20 people completed the survey and provided feedback about the wording of questions and challenges. As a result, we: \begin{itemize}
    \item Added six additional team challenges
    \item Added three additional client challenges
    \item Polished the wordings or added clarifications to decrease ambiguities
    \item Added a condition that respondents were only asked about the criticality of a challenge if they reported the challenge's frequency to be more than "Never"
    \item Added a request to share the survey to colleagues and friends working in software engineering teams
\end{itemize}

The target population of the survey were people working in software engineering without any restrictions to their role. The main survey was distributed via three different channels that looked promising in regard to response rates. The sampling method was convenience sampling. We used different survey links for each channel to get distinct response statistics. Our goal was to reach a culturally diverse set of people working in software engineering, as to avoid too many people sharing the same human value set. The targets of the channels were:

\begin{enumerate}
    \item The professional network of the research group and QualityMinds' business network, consisting of software engineers working in a variety of domains such as automotive, the financial industry and tool manufacturers (Chn1).
    \item Platforms that offer to present surveys to their users in exchange for filling out surveys of other users (Chn2): SurveyCircle.com, SurveySwap.io and SwapSurvey.com. 
    \item Uncontrolled propagation via Social Media (Chn3). We posted this survey link on Twitter, Facebook and chat communities. Additionally, we sent it to software engineers sampled from the LinkedIn search.
\end{enumerate}

The survey started on 28.08.2019 and was stopped on 24.11.2019, thus running for almost three months. We monitored how many people clicked the link and how many people started (i.e. they began filling out information) and how many people completed the survey. We then removed any answers that were of low quality. An answer was considered low quality if either the respondent wrote in the feedback field that they filled out the survey carelessly,  if any section of the team or client challenges had half or more items unanswered, or if they either failed the quality check provided with the PVQ-57RR: as it has several questions worded differently for each dimension it measures, it can detect implausible responses if the responses contradict themselves.

136 (about 70\%) of the respondents additionally left their email address to be notified about the research results, which further strengthens our confidence in the relevance of the research direction for professionals.

The most unsatisfactory approach was the survey exchange platforms, which had many response sets that were skipped through without providing answers to the questions (see Table \ref{tableResponseStats}). LinkedIn worked very well with 149 out of 166 response sets suitable for inclusion. Some people showed great interest and contacted us to reiterate the importance of research on human aspects in software engineering even though we "cold called" them.

\captionof{table}{Response Statistics}
\small
\begin{center}
\begin{xtabular}{lcccc}
  \midrule
  No. & Clicked & Started & Completed & Included  \\ [0.5ex]
  \midrule
  Chn1 & 67 & 41 & 24 & 24 \\*
  Chn2 & 49 & 32 & 28 & 19 \\*
  Chn3 & 326 & 232 & 166 & 149 \\*
  \midrule
  Total & 442 & 305 & 218 & 192 \\*
\end{xtabular}
\label{tableResponseStats}
\end{center}
\section{Study Results}\label{sec_results}
In the following, we report on the results of the survey: the context of the respondents, the frequency and criticality of our initial challenge set, whether there exists a link between human values and challenge occurrence, and the link between national diversity in the team and challenge occurrences. We conclude the report of the survey results by analysing the impact of team virtualization. Finally, we introduce the revised challenge catalogue based on the feedback from the survey.

\subsection{Demographics}\label{subDemo}
\begin{figure}
  \includegraphics[width=1\linewidth]{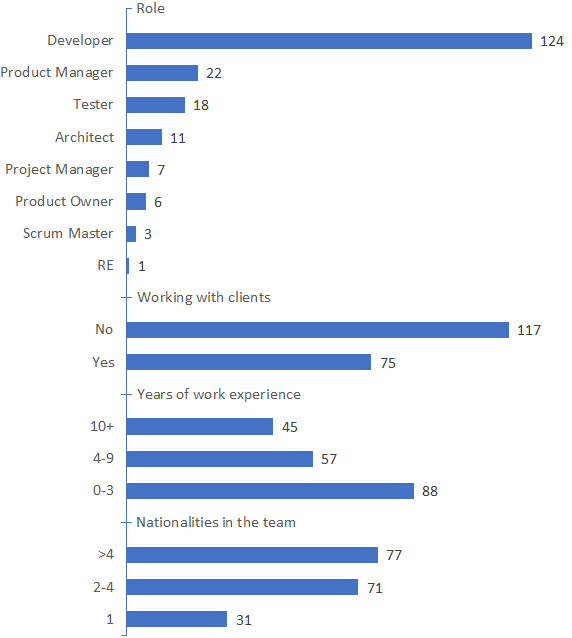}
  \caption{Respondent characteristics.}
  \label{fig:roledis}
\end{figure}

To better understand the context of the respondents, we asked them to report on their role in the team; either by selecting one of predefined roles (see Fig. \ref{fig:roledis}) or by describing it in a text field. Most of the respondents reported to be mainly a developer. 28 respondents used the comment field to explain that they assume multiple roles depending on the current need.

\begin{figure}
  \includegraphics[width=1\linewidth]{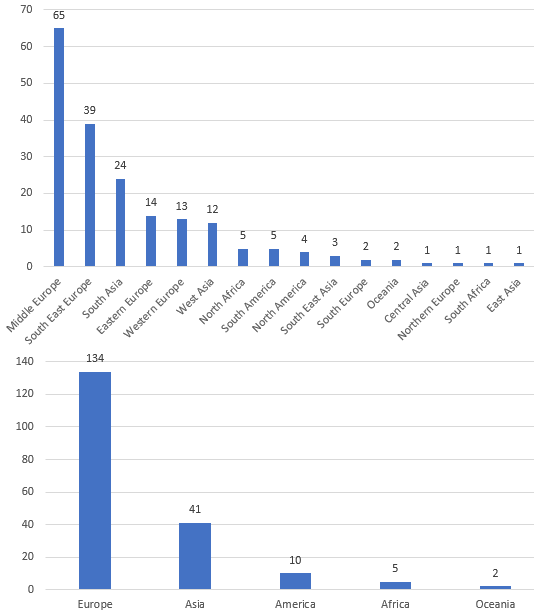}
  \caption{Nationality Distribution of the respondents, binned in regions and continents.}
  \label{fig:nationbincomb}
\end{figure}

Slightly more than half of the respondents had been working less than 5 years in the industry. Around 20\% replied that they had been working in software engineering for more than 10 years. Around 40\% of the respondents reported to have direct interaction with clients.
We also asked participants to report on their nationality as well as the nationalities present in their teams [RQ3]. In total, our respondents had 49 different nationalities. 19 of them had a dual citizenship. Our sample contained predominantly European nationalities, with 134 (69.8\%) European nationals, 41 (21.4\%) Asian nationals, and 17 (8.9\%) respondents from other regions (see Fig. \ref{fig:nationbincomb}). 145 respondents (75.5\%) worked in Germany at the time of the survey. 31 respondents were working in teams with only a single nationality present. Most respondents reported to work in very multinational teams, having five or more nationalities. 

\subsection{Challenge Evaluation}\label{subProbEval}

In order to understand which challenges matter most and to answer RQ1, we asked respondents to rate the criticality (on a 4-point scale) and frequency (on a 5-point scale) of a set of challenges. We mapped the scales to values of 0-3 for Criticality (\emph{Not problematic} = 0, \emph{Somewhat problematic} = 1, \emph{Problematic = 2}, \emph{Highly problematic} = 3) and 0-4 for Frequency (\emph{Never} = 0, \emph{Rarely} = 1, \emph{Sometimes} = 2, \emph{Often} = 3, \emph{Very often} = 4), and then calculated the mean and median of all responses for each challenge. To rank the challenges, we introduced the property \emph{Impact}, which is defined as the product of the normalized frequency and normalized criticality to remove the different weighting introduced by different scales: a challenge that occurs often and is critical has a high impact. For the team challenges, we also asked whether there were mitigation measures in place to prevent the challenge. By far the most impactful team challenge was \emph{Insufficient analysis at the beginning of a task}, followed by \emph{Lack of leadership}. From there on, impact scores were close together. Least important was \emph{People crying in discussions}, followed by \emph{Harassment} and \emph{Missing acceptance of alternative lifestyles}. Correlation between Frequency and Criticality was 0.48. An overview of the team challenge results can be found in Table \ref{teamProblemTable}.

For \emph{Insufficient analysis at the beginning of a task}, respondents reported the following mitigation measures:
\begin{itemize}
    \item Dedicated meetings for reviews, retrospection, as well as "pre-grooming" that allow special time slots for team members to ask about areas they have insufficient knowledge about
    \item Introduction of formal specification processes, to provide a framework to specify a minimum amount of information to task assignees
    \item Team commitment to clean planning, including intense communication with all stakeholders from the start
    \item Agile approaches such as SCRUM
\end{itemize}

For \emph{Lack of leadership}, respondents reported the following:
\begin{itemize}
    \item Coachings for personnel in leading roles, as well as leaders-to-be
    \item Fostering of leaders in-house with senior guidance
    \item Dedicated feedback sessions and retrospectives
    \item The employment of software to measure office/team mood, and frequent supervision of these reports by the leader
\end{itemize}

A list of aggregated mitigation measures for each team problem can be found in Appendix \ref{appendix_miti}, while the full data about reported mitigations can be found browsing the link to the data repository in Appendix \ref{appendix_repo}.

\hfill \break

For the client challenges, the most impactful challenge was \emph{Lack of interest in the project by the client}, \emph{Client does not know what they want} and \emph{No direct communication with client}. Client challenges with the least impact were \emph{Misinterpretation} and \emph{Lack of communication}. The correlation between Frequency and Criticality was 0.46. All results can be found in Table \ref{clientprobtable}.

\begin{table*}[t]
\captionof{table}{Team Problems, ranked by Impact. $\pm$ denotes standard deviation. Mitigation refers to the percentage of respondents that reported any mitigation measure to support this challenge was in place.}
\small
  \centering
\begin{tabular}{llcccccc}
 \midrule
No. & Problem & \multicolumn{2}{c|}{Frequency} & \multicolumn{2}{c|}{Criticality} & Impact & Mitigated \\ [0.5ex]
 &  & Mean & Median & Mean & Median &  &  \% \\
 \midrule
T13 & Insufficient analysis at the beginning of a task & 2.43$\pm$1.02 & 2 & 2.07$\pm$0.84& 2 & 0.252 & 0.463 \\
T4 & Lack of leadership & 2.21$\pm$1.11 & 2 & 1.91$\pm$0.95& 2 & 0.210 & 0.389\\
T18 & Missing documentation of the project & 2.59$\pm$1.06 & 3 & 1.59$\pm$0.83& 2 & 0.206 & 0.489\\
T7 & Demotivation & 2.04$\pm$1.17 & 2 & 1.92$\pm$0.97& 2 & 0.196 & 0.368\\
T19 & Information is not made known to the team & 2.06$\pm$1.12 & 2 & 1.86$\pm$0.94& 2 & 0.192 & 0.384\\
T3 & Lack of teamwork skills & 1.98$\pm$1.03 & 2 & 1.89$\pm$0.95& 2 & 0.187 & 0.459\\
T20 & Misunderstandings in communication & 2.05$\pm$1.01 & 2 & 1.82$\pm$0.89& 2 & 0.186 & 0.392\\
T5 & Communication plan is neglected & 2.15$\pm$1.09 & 2 & 1.73$\pm$0.89& 2 & 0.185 & 0.407\\
T12 & Insufficient collaboration & 1.87$\pm$1.11 & 2 & 1.97$\pm$0.95& 2 & 0.184 & 0.475\\
T14 & Subjective interpretations of tasks & 2.24$\pm$1.09 & 2 & 1.63$\pm$0.88& 2 & 0.183 & 0.328\\
T25 & Slow decision making & 1.95$\pm$1.17 & 2 & 1.80$\pm$0.87& 2 & 0.176 & 0.258\\
T1 & Lack of experience & 2.29$\pm$0.98 & 2 & 1.51$\pm$0.87 & 2& 0.172 & 0.568\\
T17 & Conflicts of interests at management level & 1.77$\pm$1.26 & 2 & 1.85$\pm$0.97& 2 & 0.164 & 0.133\\
T30 & People not reporting problems in time & 1.68$\pm$1.06 & 2 & 1.93$\pm$0.89& 2 & 0.162 & 0.331\\
T21 & Certain people dominating discussions & 2.24$\pm$1.17 & 2 & 1.40$\pm$0.97& 1 & 0.157 & 0.210\\
T15 & Work is not solution-oriented & 1.78$\pm$1.10 & 2& 1.73$\pm$0.95& 2 & 0.154 & 0.383\\
T8 & Gold plating & 1.98$\pm$1.14 & 2 & 1.49$\pm$1.05& 1 & 0.148 & 0.380\\
T6 & Business needs are neglected & 1.64$\pm$1.08 & 2 & 1.76$\pm$0.92& 2 & 0.145 & 0.372\\
T2 & Lack of qualification & 1.82$\pm$1.05 & 2& 1.59$\pm$0.93 & 2 & 0.144 & 0.581\\
T31 & Over-Confidence & 1.83$\pm$1.03 & 2 & 1.54$\pm$0.94 & 2& 0.141 & 0.157\\
T11 & Missing willingness to adapt to new processes & 1.85$\pm$1.12 & 2 & 1.47$\pm$0.88& 1 & 0.136 & 0.307\\
T9 & Missing willingness to adapt to new software & 1.71$\pm$1.07 & 2 & 1.37$\pm$0.85& 1 & 0.117 & 0.268\\
T32 & Pressure from client being forwarded to SE team & 1.68$\pm$1.13 & 2 & 1.39$\pm$0.89& 1 & 0.116 & 0.316\\
T10 & Missing willingness to adapt to new infrastructure & 1.68$\pm$1.13 & 2 & 1.35$\pm$0.86& 1 & 0.114 & 0.298\\
T22 & Missing respect in the workplace & 1.14$\pm$1.11 & 1 & 1.66$\pm$1.06& 2 & 0.095 & 0.371\\
T29 & Frequently changing team members & 1.17$\pm$1.10 & 1 & 1.62$\pm$0.92& 2 & 0.094 & 0.266\\
T26 & Conflicting working styles & 1.72$\pm$1.03 & 2 & 1.09$\pm$0.83& 1 & 0.094 & 0.227\\
T27 & People getting angry in discussions & 1.28$\pm$1.00 & 1 & 1.38$\pm$1.03 & 1& 0.088 & 0.301\\
T33 & Exaggerated seeking of project problems & 1.32$\pm$1.04 & 1 & 1.31$\pm$0.87& 1 & 0.086 & 0.182\\
T16 & Language barriers & 1.19$\pm$1.02 & 1 & 1.20$\pm$0.85& 1 & 0.071 & 0.427\\
T23 & Missing acceptance of alternative lifestyles & 0.93$\pm$1.14 & 1 & 1.21$\pm$0.99& 1 & 0.056 & 0.339\\
T24 & Harassment & 0.37$\pm$0.72 & 0 & 1.58$\pm$1.22 & 2& 0.029 & 0.474\\
T28 & People crying in discussions & 0.23$\pm$0.57 & 0 & 1.20$\pm$1.17& 1 & 0.014 & 0.198\\
\end{tabular}
\label{teamProblemTable}
\end{table*}

\begin{figure}
  \includegraphics[width=1\linewidth]{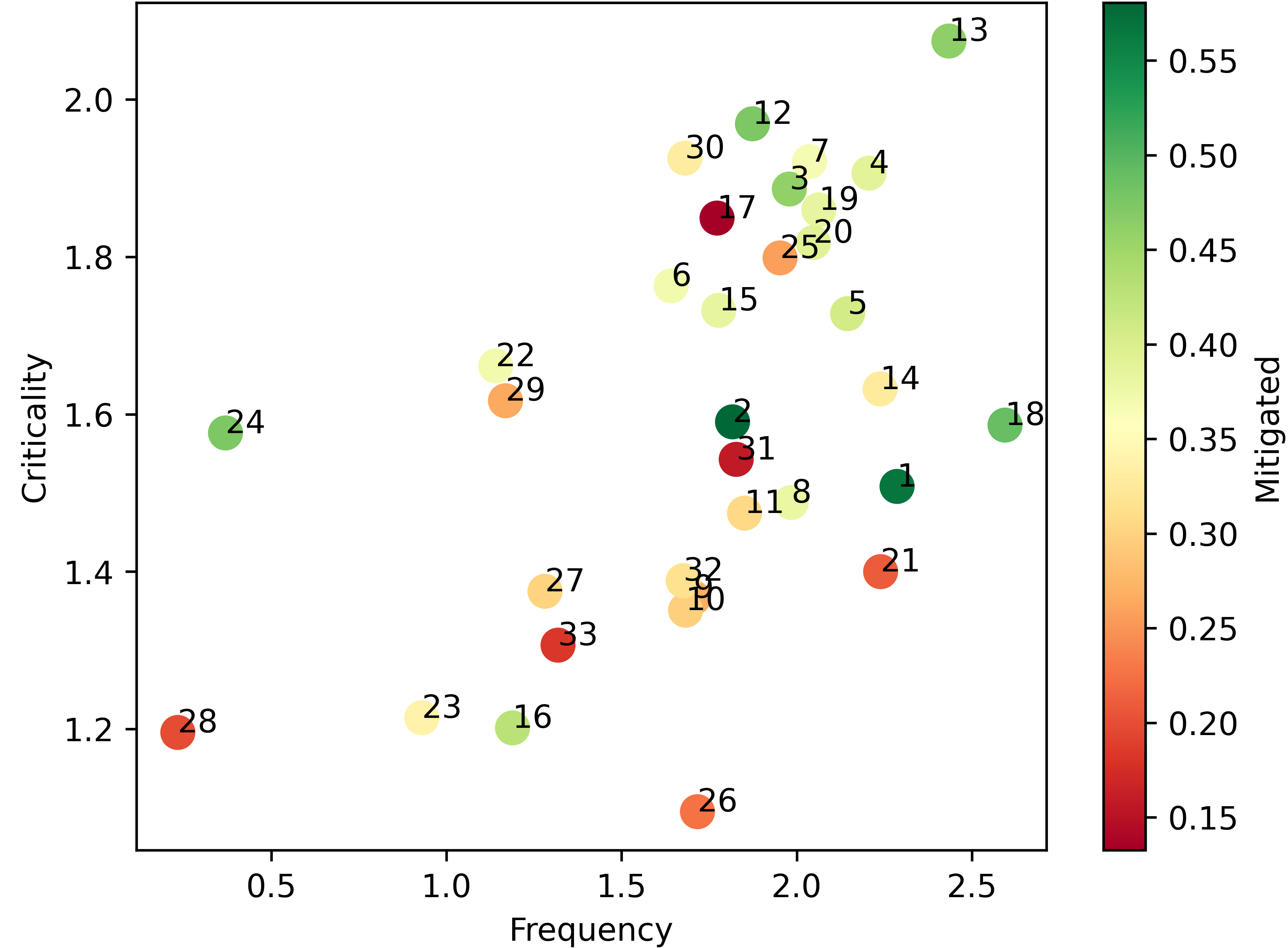}
  \caption{Team challenges, displaying Frequency on the horizontal axis, Criticality on the vertical axis and percentage of respondents having mitigation measures in place in the sample as color coded.}
  \label{fig:teamprobfig}
\end{figure}

\begin{table*}[t]
\small
\captionof{table}{Client Problems, ranked by Impact. ($\pm$ denotes standard deviation)}
  \centering
\begin{tabular}{llcccccc}
 \midrule
No. & Problem & \multicolumn{2}{c|}{Frequency} & \multicolumn{2}{c|}{Criticality}& Impact \\ [0.5ex]
 &  & Mean & Median & Mean & Median &  \\
\midrule
C3 & Lack of interest in the project by the client & 2.52$\pm$0.96 & 2 & 2.17$\pm$0.84 & 2 & 0.273\\
C2 & Client does not know what they want & 2.26$\pm$0.99 & 2 & 2.28$\pm$0.83 & 2 & 0.258\\
C15 & No direct communication with client & 2.32$\pm$0.93 & 2 & 1.97$\pm$0.85 & 2 & 0.228\\
C9 & Client unable to specify measurable non-functional reqs & 2.33$\pm$1.09 & 2 & 1.85	$\pm$0.87 & 2 & 0.216\\
C14 & Insufficient analysis at the beginning of the project & 2.00$\pm$1.04 & 2 & 1.92$\pm$0.94 & 2 & 0.192\\
C25 & Over-Confidence & 2.06$\pm$1.00 & 2 & 1.87$\pm$0.89 & 2 & 0.192\\
C24 & People not reporting problems in time & 2.17$\pm$1.16 & 2 & 1.77$\pm$0.91 & 2 & 0.192\\
C21 & Missing respect towards the client from the team & 2.04$\pm$0.95 & 2 & 1.85$\pm$0.85 & 2 & 0.189\\
C10 & Unclear roles and responsibilities at client side & 2.29$\pm$1.13 & 2 & 1.60$\pm$0.86 & 2 & 0.183\\
C5 & Missing technical knowledge at client side & 2.34$\pm$1.21 & 2.5 & 1.51$\pm$0.98 & 1 & 0.177\\
C8 & Client unable to specify functional requirements & 2.01$\pm$1.04 & 2 & 1.70$\pm$0.98 & 2 & 0.171\\
C12 & Weak management at client side & 2.19$\pm$1.13 & 2 & 1.56$\pm$0.84 & 2 & 0.171\\
C7 & Conflicts of interests at client side & 2.07$\pm$1.05 & 2 & 1.57$\pm$1.02 & 2 & 0.163\\
C6 & Exaggerated quality expectation of the client & 2.53$\pm$1.16 & 3 & 1.28$\pm$0.88 & 1 & 0.162\\
C11 & Lack of prioritization by client & 2.01$\pm$1.14 & 2 & 1.53$\pm$0.93 & 2 & 0.154\\
C17 & Subjective interpretations of tasks & 1.88$\pm$0.96 & 2 & 1.63$\pm$0.83 & 2 & 0.153\\
C13 & Insufficient collaboration & 2.04$\pm$1.16 & 2 & 1.46$\pm$0.99 & 2 & 0.149\\
C18 & Work is not solution-oriented & 1.91$\pm$1.02 & 2 & 1.55$\pm$0.85 & 2 & 0.148\\
C4 & Missing IT project experience at client side & 1.58$\pm$1.02 & 2 & 1.72$\pm$1.01 & 2 & 0.136\\
C27 & Exaggerated seeking of project problems & 1.65$\pm$1.34 & 2 & 1.51$\pm$0.91 & 1 & 0.124\\
C26 & Management restricting communication of problems to client & 1.78$\pm$1.12 & 2 & 1.38$\pm$0.95 & 1 & 0.123\\
C19 & Language barriers & 1.53$\pm$1.07 & 1 & 1.58$\pm$0.97 & 2 & 0.121\\
C16 & Communication plan is neglected & 1.29$\pm$1.01 & 1 & 1.59$\pm$0.98 & 2 & 0.102\\
C23 & Delays due to dependencies to client's third parties & 1.26$\pm$1.04 & 1 & 1.60$\pm$0.96 & 2 & 0.100\\
C22 & Missing respect towards the team from the client & 1.19$\pm$1.10 & 1 & 1.52$\pm$1.01 & 1.5 & 0.091\\
C1 & Lack of communication & 1.32$\pm$1.04 & 1 & 1.31$\pm$0.87 & 1 & 0.086\\
C20 & Misinterpretations & 1.00$\pm$0.96 & 1 & 1.28$\pm$0.93 & 1 & 0.064\\
\end{tabular}
\label{clientprobtable}
\end{table*}

\begin{figure}
  \includegraphics[width=1\linewidth]{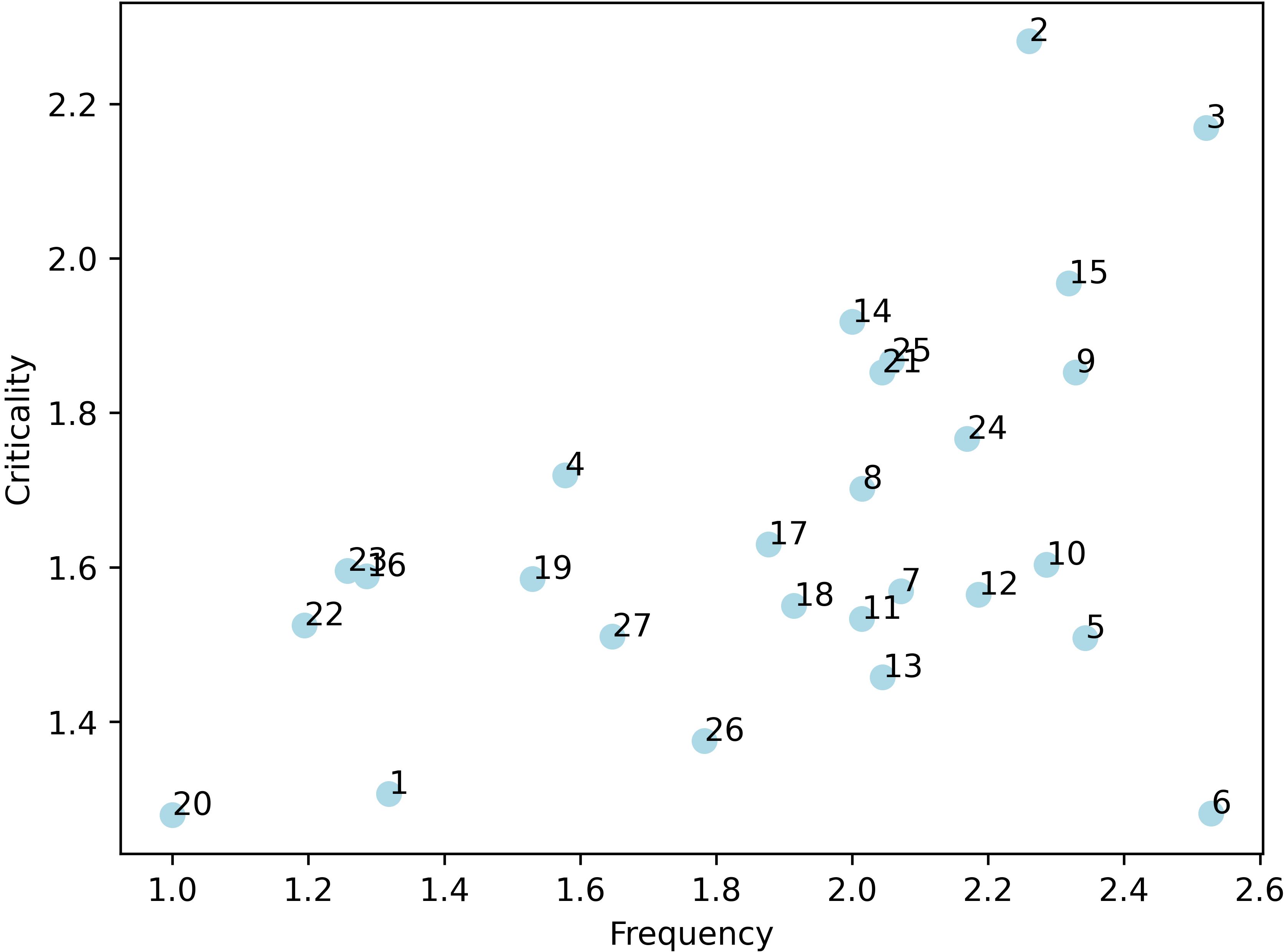}
  \caption{Client challenges, displaying Frequency on the horizontal axis and Criticality on the vertical axis.}
  \label{fig:clientprobfig}
\end{figure}
\subsection{Personal Values}\label{subValDiv}
RQ2 aimed at exploring whether the values of an individual had any impact on the reported frequency or criticality on the challenges of our challenge set. For use with correlation analyses, Schwartz describes steps to calculate \emph{centered human values} \cite{schwartz2012refining} for the 19 human value dimensions (see Section \ref{subsub_persval}): First, the mean of the items for the respective value dimensions is taken. Then, the mean of all items is calculated. Finally, the overall mean is subtracted from the mean of each dimension. We then calculated the Pearson correlation coefficient between the human value dimensions and the frequency and criticality of each challenge. All correlations were weak with no correlation concerning the frequency exceeding 0.27, and no correlation concerning the criticality exceeding 0.34. Additionally, as correlating 33 challenges with 19 human values results in many (627) tests, we applied the Holm–Bonferroni method\cite{holm1979} to adjust the p-values and found none to be statistically significant (alpha=0.05).

We then aggregated the 19 values into the four "higher order" values \emph{Self-Transcendence, Self-Enhancement, Openness to change and Conservation} that were also described by Schwartz \cite{schwartz2012refining}, but the resulting correlations were even weaker (no frequency correlation exceeding 0.22 and no criticality correlation exceeding 0.26), none of which were statistically significant.

Therefore and to answer RQ2, we report that we found no evidence regarding the link between an individual's human values and the occurrence of human challenges in software engineering teams. 
\subsection{Nation Count in Teams}\label{subNatDiv}

We additionally examined whether the presence of multiple nationalities in the team had an influence on the frequency and criticality of the team challenges.

\begin{figure}[ht]
  \includegraphics[width=1\linewidth]{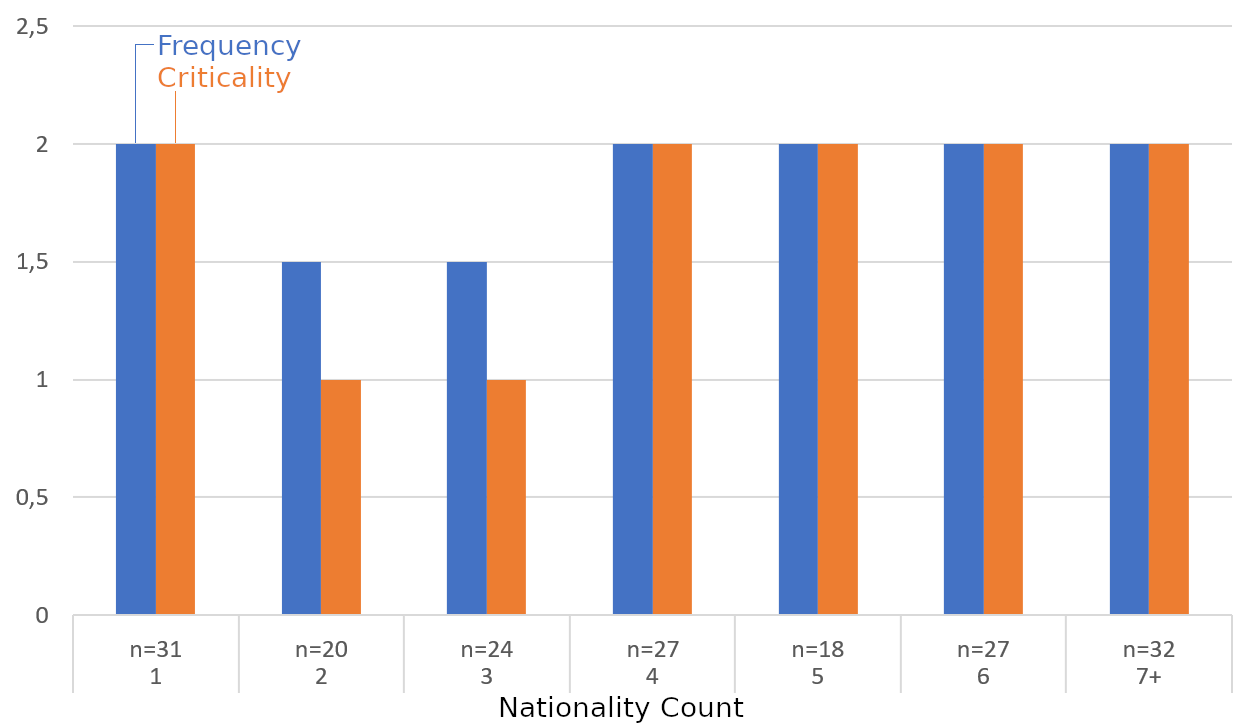}
  \caption{The median scores of all challenges relative to the nationality count in the team. For nation counts of two and three, Frequency drops by 0.5, for Criticality, it drops a whole level, from 2 (\emph{Problematic}) to 1 (\emph{Somewhat problematic}).}
  \label{natCountFig}
\end{figure}

31 respondents worked in teams having just one nationality, while 148 worked in teams with more than one nationality. 13 respondents did not answer this question. For each nation count, we calculated the median of all challenge medians of the respondents that reported this nation count for both Frequency and Criticality up until a nation count of 6. For a nation count of 7 and higher, there were very few respondents for each count, so those were binned. As shown in Fig. \ref{natCountFig}, there was a drop in reported Frequency and Criticality for a nation count of 2 and 3.

We used this insight to create three bins: A bin with respondents having a single-nationality team (n=31), a bin with 2-3 nationalities (n=44), and a bin with four and more nationalities (n=104). 

First, we compared the 2-3 group with the single nationality group and checked for statistically significant (alpha=0.05) differences to single challenges. None of the challenges were significantly more critical or more frequent in the 2-3-group. However, the following challenges were significantly more frequent for the single-nationality teams:

\begin{itemize}
    \item T3 \emph{Lack of teamwork skills}\newline(Freq: +0.5883, p=0.0149)
    \item T12 \emph{Insufficient collaboration}\newline(Freq: +0.5652, p=0.0405)
    \item T23 \emph{Missing acceptance of alternative lifestyles}\newline (Freq: +0.7394, p=0.0115)
\end{itemize}

The following challenges were significantly more critical:
\begin{itemize}
    \item T3 \emph{Lack of teamwork skills}\newline(Crit: +0.6656, p=0.0112)
    \item T9 \emph{Missing willingness to adapt to new software}\newline(Crit: +0.6923, p=0.0031)
    \item T10 \emph{Missing willingness to adapt to new infrastructure}\newline(Crit: +0.7495, p=0.0008)
    \item T11 \emph{Missing willingness to adapt to new processes}\newline(Crit: +0.735, p=0.002)
    \item T12 \emph{Insufficient collaboration}\newline(Crit: +0.5733, p=0.0338)
    \item T15 \emph{Work is not solution-oriented}\newline(Crit: +0.6494, p=0.0168)
    \item T17 \emph{Conflicts of interest at management level}\newline(Crit: +0.6786, p=0.0236)
    \item T28 \emph{People crying in discussions}\newline(Crit: +1.0714, p=0.008)
    \item T33 \emph{Exaggerated seeking of project problems}\newline(Crit: +0.6737, p=0.0076)
\end{itemize}

Secondly, we compared the 2-3 group with the group having four or more nationalities. Again, no challenge was significantly more critical or more frequent for the 2-3-group. The following challenges were significantly more frequent for the 4+ group:

\begin{itemize}
    \item T2 \emph{Lack of qualification}\newline(Freq: +0.5541, p=0.0025)
    \item T3 \emph{Lack of teamwork skills}\newline(Freq: +0.6986, p=0.0001)
    \item T12 \emph{Insufficient collaboration}\newline(Freq: +0.4126, p=0.0294)
    \item T29 \emph{Frequently changing team members}\newline(Freq: +0.3831, p=0.0483)
    \item T31 \emph{Over-Confidence}\newline(Freq: +0.3672, p=0.0434)
\end{itemize}

The following challenges were significantly more critical for the 4+ group:
\begin{itemize}
    \item T1 \emph{Lack of experience}\newline(Crit: +0.3311, p=0.0363)
    \item T2 \emph{Lack of qualification}\newline(Crit: +0.4, p=0.0243)
    \item T3 \emph{Lack of teamwork skills}\newline(Crit: +0.6595, p=0.0005)
    \item T4 \emph{Lack of leadership}\newline(Crit: +0.3652, p=0.0362)
    \item T5 \emph{Communication plan is neglected}\newline(Crit: +0.3749, p=0.0365)
    \item T7 \emph{Demotivation}\newline(Crit: +0.3904, p=0.0354)
    \item T8 \emph{Gold plating}\newline(Crit: +0.7778, p=0.0001)
    \item T9 \emph{Missing willingness to adapt to new software}\newline(Crit: +0.4337, p=0.0103)
    \item T10 \emph{Missing willingness to adapt to new infrastructure}\newline(Crit: +0.5016, p=0.0026)
    \item T11 \emph{Missing willingness to adapt to new processes}\newline(Crit: +0.447, p=0.0063)
    \item T12 \emph{Insufficient collaboration}\newline(Crit: +0.3986, p=0.0347)
    \item T24 \emph{Harassment}\newline(Crit: +1.0667, p=0.0039)
    \item T28 \emph{People crying in discussions}\newline(Crit: +1.2143, p=0.0029)
\end{itemize}

This finding indicates there seems to be a "sweet spot" of nationality diversity in our data. However, the count alone is a very raw dimension with limited explanatory power. We believe further investigation is necessary to understand these effects. To answer RQ3, there seems to be evidence that the degree of multinationality has an impact on the frequency as well as the criticality of some challenges.
\subsection{Team Virtuality}\label{subTeamVirt}

\begin{figure}[ht]
  \includegraphics[width=1\linewidth]{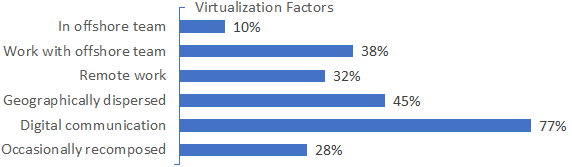}
  \caption{Team Virtualization characteristics of the survey respondents.}
  \label{figTV}
\end{figure}

We asked six questions to assess the work contexts of respondents in regard to degree of team virtualization (see Table \ref{demotable} and Fig.  \ref{figTV}). Each question could either be answered with "Yes" or "No".

To answer RQ4, we split the sample for each virtualization characteristic and performed an independent t-test for each team challenge, using an alpha of 0.05 as a cutoff value for the p-values. The data looked normally distributed for all criticality values and for most frequency items, which was confirmed by a Shapiro-Wilk test \cite{SHAPIRO1965}. The use of a t-test is acceptable for data that is not normally distributed according to the Central Limit Theorem if the sample size is sufficient (\textgreater30 per group) \cite{bortz2010tests}. For those challenge frequencies the Shapiro-Wilk test returned an ambiguous outcome to (T1, T21, T22, T27), we checked whether there was a difference in statistical significance using a Mann–Whitney U test \cite{mann1947test} instead of a t-test. This would make T27 statistically significant regarding increased frequency for \emph{Working with an offshore team} (p=0.0479) and \emph{Working in a geographically dispersed team} (p=0.0497). However, given that they do not show up in the t-test and the p value is very close to the threshold of 0.05, we chose to omit these occurrences.
\newline

Being \textbf{part of an offshore team} does not seem to have an effect on the frequency of any challenge. However, T26 "\textit{Conflicting working styles}" is regarded as less critical (Crit: -0.496, p=0.0319). It is noteworthy that this is the only case where virtualization had a positive impact on a challenge according to the data.

\textbf{Working with an offshore team} was associated with a small, but significant increase in the frequency of the challenges 

\begin{itemize}
    \item T6 \emph{Business needs are neglected}\newline(Freq: +0.342, p=0.0376)
    \item T29 \emph{Frequently changing team members}\newline(Freq: +0.350, p=0.0328)
\end{itemize}
It did not have a statistically significant effect on the criticality of any challenge.

\textbf{Working from remote} also led to a small increase of T29 (Freq: +0.366, p=0.0307). There was no significant effect on the criticality of any challenge.

Working in a \textbf{geographically dispersed team} was associated with an increase of the frequency of 
\begin{itemize}
    \item T6 \emph{Business needs are neglected}\newline(Freq: +0.405, p=0.0111)
    \item T17 \emph{Conflicts of interest at management level}\newline(Freq: +0.418, p=0.0231)
    \item T32 \emph{Pressure from client forwarded to SE team level}\newline(Freq: +0.397, p=0.0181)
    \item T33 \emph{Exaggerated seeking of project challenges level}\newline(Freq: +0.372, p=0.0016)
\end{itemize}
Additionally, T1 "\textit{Lack of experience}" (Crit: +0.285, p=0.0271) was considered more critical.

Using \textbf{digital communication} led to more frequent occurrence of
\begin{itemize}
    \item T16 \emph{Language Barriers}\newline(Freq: +0.422, p=0.0161)
    \item T29 \emph{Frequently changing team members}\newline(Freq: +0.458, p=0.0159)
     \item T32 \emph{Insufficient collaboration}\newline(Freq: +0.398, p=0.0437)
\end{itemize}
There was no significant effect on criticality.

Finally, working in \textbf{teams that are occasionally recomposed} was associated with an increase in the frequency of

\begin{itemize}
    \item T15 \emph{Work is not solution-oriented} \newline(Freq: +0.366, p=0.0408)
    \item T22 \emph{Missing respect in the workplace} \newline(Freq: +0.634, p=0.0398)
     \item T24 \emph{Harassment}\newline(Freq: +0.220, p=0.0176)
     \item T25 \emph{Slow decision making}\newline(Freq: +0.404, p=0.0322)
     \item T27 \emph{People getting angry in discussions}\newline(Freq: +0.396, p=0.0132)
     \item T29 \emph{Frequently changing team members}\newline(Freq: +0.929, p=0.0001)
     \item T30 \emph{People not reporting problems in time}\newline(Freq: +0.368, p=0.0305)
\end{itemize}

No statistically significant effect on criticality was observed.

To answer RQ4, yes, the degree of virtualization seems to have an impact on human challenges in the team. Increasing the degree of virtualization of a team may increase the frequency of some human challenges, but only affected the criticality for one challenge and only for geographically dispersed teams. 
\subsection{Challenge Catalogue}\label{subNewProbCat}
In the survey, in addition to checking frequency and criticality on an initial set of challenges, we collected feedback in order to create a relevant, up-to-date contemporary set of problems, ranging from suggestions to improve the wordings or reduce ambiguity as well as additional challenges that respondents felt were not covered by the survey challenge set. The survey had three optional open text fields at the end to collect feedback: Additional team challenges, additional client challenges, and general feedback about the survey. 74 respondents contributed feedback about additional team challenges that they felt were not covered sufficiently. 

The feedback was analyzed by a member of the research group and split into challenges; either existing ones from the set if applicable or new ones if the match did not seem close enough. Then, a second colleague was tasked to match the feedback to the combined set of initial challenge codes plus the new ones discovered by the first member. In a second step, these challenges were aggregated or discarded if they did not seem to address the human side of the challenge well enough, or they were too close to existing challenges and thus redundant. The detailed list of changes to the initial challenge set can be found in the public repository (see Appendix \ref{appendix_repo}); the revised challenges are shown in Table \ref{newProbTable}.

Respondents also reported that the existing challenges were in part ambiguous. Some challenges were very specific, while others were more abstract. For our revised challenge catalogue, we aggregated challenges and leveled the abstraction grade. We removed challenges that are difficult to diagnose, are close to being an organizational matter rather than a human challenge, or do not affect the actual software engineering work. Additionally, we plan to better describe how that kind of challenge might express itself in practise. For future surveys, we hope to reduce misinterpretations of the challenges, reduce survey length and be able to more exactly map the state of human challenges in software engineering. To limit the scope, our refined set of challenges focuses on challenges within the company (Team Challenges) and omits challenges happening in the context between the company and external entities (Client Challenges).

Aggregating the challenges, we split the challenges into two categories (see Table \ref{newProbTable}): 
\begin{itemize}
    \item Interpersonal. This category contains challenges that occur when people interact with each other. These challenges happen between colleagues (and are witnessed by the challenge reporter) or from colleagues to the respondent. They include problematic demeanour, clashes in the way people communicate, or general stances of individuals that prevent collaboration.
    \item Intrapersonal. This category contains burdens of any sort on an individual that may impact the work they do in a negative way. This category comprises all reported challenges that are an emotional state of an individual, or a stance they have towards a project, way of working or process.
\end{itemize}
\captionof{table}{Revised Team Challenge Set}
\small
\begin{center}
\label{newProbTable}
\begin{xtabular}{lp{.335\textwidth}}
  \midrule
  No. & Challenge  \\
  \midrule
  & Inter-Personal Challenges\\
  \midrule
    1 & “Not my job”-mentality\\
    2 & Different communication preferences\\
    3 & Destructive Competition\\
    4 & Lack of Confidence\\
    5 & Over-Confidence\\
    6 & Lack of Empathy\\
    7 & Pressure\\
    8 & People not reporting problems in time\\
    9 & Frequently changing team members\\
    10 & Aggressive Discussion Styles\\
    11 & Misunderstandings\\
    12 & Discussions are not solution-oriented\\
    13 & Reluctance to change working habits\\
    14 & Team is not kept up-to-date\\
    15 & Lack of mentors\\
    16 & Missing team cohesion\\
    17 & Different needs of staff not respected\\
    18 & Negativity\\
    19 & Harassment\\
    20 & Missing appreciation\\
    21 & Lack of mediation in conflicts\\
  \midrule
     & Intra-Personal Challenges\\
  \midrule
    1 & Bad Work-Life-Balance\\
    2 & Work is not solution-oriented\\
    3 & Tasks insufficiently analyzed\\
    4 & Task is not well defined or explained\\
    5 & Lack of knowledge\\
    6 & Gold Plating\\
    7 & Demotivation\\
\end{xtabular}
\end{center}  

The categorization allows to examine how certain measures at work affect an area of a challenge as a whole and puts some of the challenges into more specific context. If a measure is expected to only affect one area of challenges, the unnecessary other category can also be excluded to make surveys more compact. 

A list of descriptions of the new challenge set can be found online in the repository linked in appendix \ref{appendix_repo}. By explaining each problem, we additionally hope to decrease misunderstandings and conveying the scope of each individual challenge.

\section{Discussion}\label{sec_conclusion}
In this article, we summarised our findings from our survey conducted among 192 software development professionals. Our research focus was finding out which human challenges are considered most frequent and critical in software engineering. We examined the links between team virtualization, human values and nationalities and the occurrence and reported criticality of challenges.

Based on the feedback of the survey, we refined the challenge set and proposed two categories.

Our initial research questions were answered as followed:
\begin{itemize}
    \item \textbf{RQ1}. We evaluated our initial challenge set. (see Table \ref{teamProblemTable}). We introduced a refined challenge set based on the feedback of the survey (See Table \ref{newProbTable}). 
    \item \textbf{RQ2}. We found no significant evidence that human values as defined by Schwartz \cite{schwartz2012refining} have an impact on the reported challenge frequency and criticality.
    \item \textbf{RQ3}. We found evidence that there seems to be a connection between nationality and challenges, with the sweet spot being teams with 2-4 nationalities.
    \item \textbf{RQ4}. We found evidence that a higher degree of team virtualization leads to a higher frequency of some challenges.
\end{itemize}

For the team challenges, \emph{Insufficient analysis at the beginning of a task} was the most critical challenge. This makes sense as it can lead to multiple consecutive problems: Time, staff, and resources may not be correctly allocated. Required expertise might be missing in the team. Costs can explode beyond assumptions. However, it is not yet clear what causes this challenge. It could be a problem of the software engineering process, which should explicitly reserve time to analyse tasks before work on it begins. It could also be due to people not voicing concerns or not having the opportunity to do so. To tackle this, it should be ensured there is an explicit time and place to discuss tasks in depth with stakeholders and people that will work on the task, to ensure everyone is synchronized well - a central tenet of scaling agile frameworks \cite{leffingwell2010agile}.

Other relevant team challenges (see Fig. \ref{fig:teamprobfig} and Table \ref{teamProblemTable}) stemmed from a multitude of causes: Own emotional states such as \emph{"Demotivation"} (T7), behaviour of superiors expressed as \emph{"Leadership"} (T4), but also synchronisation issues such as \emph{"Insufficient analysis of tasks"} (T13) or \emph{"Information not made known to the team"} (T19). None of these challenges can be linked to a common root. This indicates that human challenges are a broad and complex problems. Finding out how to support the overcoming of these challenges will likely require detailed investigation for each challenge individually. Challenges related to intolerance were among the least relevant challenges for SE professionals: \emph{"Harassment"} (T24), \emph{"People crying"} (T28) and \emph{"Missing Acceptance of alternative lifestyles"} (T23) were among the least frequent challenges. This seems to be a good sign for the state of minority inclusion in software engineering. Also, surprisingly, \emph{"Language Barriers"} (T16) are not reported to be a big challenge - while \emph{"Misunderstandings in communication"} (T20) ranked 7th in Impact. Misunderstandings thus seem to occur on a higher level than simple syntactical language mistakes. As a misunderstanding can be highly problematic, with consequences ranging from a task incorrectly implemented to a personal relationship at work being negatively affected, that certainly highlights the need for in-depth analysis of communication issues in software engineering.

The most relevant challenges between clients and SE teams (see Fig. \ref{fig:clientprobfig} and Table \ref{clientprobtable}) do seem to share a common root: Synchronisation between the client and the team is indirect and not frequent enough. The worst reported challenge was \emph{"Lack of interest in the project by the client"}. This comes as a surprise as the client is paying for the project and most likely does have an interest in getting what they envision. It could be that the feeling that there is a lack of interest, as well as the 2nd (\emph{"Client does not know what they want"} (C2)) and 3rd (\emph{"No direct communication with client"} (C15)) most relevant challenges are intertwined. The client might not be uninterested in the project, but is not aware of the process, a lack of proactivity on their side or of misunderstandings that are going on. If communication is not direct, but forwarded or potentially reworded by another team, there could be more misunderstandings. Further research should examine whether direct, frequent communication between the clients and the SE team could improve the first five relevant challenges.

We found that human values of the respondent had no significant effect on the reported frequency or criticality. We assume that the effect size is small and a larger sample size is needed. It could also be that the instrument used was inadequate for the context, and a different mapping of human values is better suited. Important cultural context may also be lost when values are reduced to numbers \cite{pudelko2015recent}. Ultimately, supporting teams or teaming up people based on human values may not be desirable in practical settings due to privacy concerns and discriminatory effects. 

Teams with 2-3 nationalities among the team members had less frequent and critical challenges than teams with just a single nation in our data. Interestingly, half of the less critical challenges have to do with "Missing willingness to adapt". Maybe the exposure to different nationalities leads to more acceptance of different stances on procedures, thus disagreement is seen as less critical to the project's success. 
However, the advantage disappeared and challenges became more frequent and critical with teams with more than five nationalities. Teams with many nationalities are probably more likely virtual teams or consist of subcontracted subteams: in our data, 76\% of the 2-3 nation counts bin was geographically dispersed, but 96\% of the 4+ nationalities bin. In this case, the reason for a higher frequency and perceived criticality of problems might be related to the organisational structure of the team rather than the fact that there are many nationalities.

Most aspects of team virtualization added to the frequency of challenges. Only one challenge was more critical: \emph{Lack of experience} in case of geographically dispersed teams. It could be that less experienced workers may need more active supervision that is hard to maintain if colleagues are not physically available. For workers in offshore teams, the criticality was less severe when it came to conflicting working styles. Maybe that is because offshore teams tend to be used to outsource implementation rather than design, so the task merely needs to be done in whatever way.

An interpretation as to why the frequency rises in cases of many virtualization aspects could be this: Remote work, geographically dispersed teams and digital communication reduce personal interactions and may lead to weaker personal bonds between colleagues or, in case of re-composition of the team, bonds that are regularly broken, thus social roles have to be re-learnt frequently (possibly contributing to T25 and T30). This could lead to less understanding of each other and the social structure of a team, in turn leading to less loyalty towards the company (contributing to T6, T29) or colleagues (T22, T24, T27).

Interestingly, the use of digital communication seemed to contribute to the frequency of language barriers. Intuitively, using text based communication should decrease the frequency as the involved can look up words and acoustic understanding is not a challenge. However, maybe bad connections – still common as of 2020 – contribute to people not understanding each other clearly, which in combination with unusual pronunciation or wordings may contribute to a feeling of the language being a barrier.

\subsection{Limitations}
Most of the response sets came from LinkedIn users. As the LinkedIn search prioritizes local people, many of the respondents (75.5\%) were working in Germany. As Europe's workforce differs greatly regarding engagement, happiness and productivity (with Germany having only few highly engaged workers) \cite{schaufeli2018work}, it can be assumed that the frequency and criticality of challenges could change if the focus would lie on a different country. The nationalities were predominantly European (69.8\%), some from Asia (21.4\%), but only few from America, Africa and other regions. Older software engineers that are less likely to use social media may be underrepresented. 24 of the responses were from members of the QualityMinds company network, which might have introduced a bias regarding the industry context, as the company at the time of the survey primarily operated in the financial sector. 

The initial challenge set came largely from the NaPiRE dataset \cite{mendez2016naming}, which had requirement engineers as a target group. Also, some respondents reported the challenges to be ambiguous or unclear. This became more obvious in the analysis of the data and the feedback: The initial catalogue contained problems that are not strictly enough a human challenge, such as "Missing documentation of the project". Some challenges were too open to interpretation or too similar - a small explanation for each challenge with examples of how it may manifest can help with getting the scope of the challenge across. The refinement of the challenge catalogue is prone to subjectivity of the two coders and represents our point of view on the matter. However, the data set is disclosed - other researchers are invited to review it.

Some of the challenges have a U-shaped relationship, i.e. depending on the subjective thresholds of the observer, reducing the symptoms of a challenge could create another challenge: A worker who wants to make an effort to reduce their perceived demotivation could end up being perceived as over-confident or indulging in gold plating. Colleagues that want to get everyone on board and establish an atmosphere of collaboration could be regarded as not working in a solution-oriented way. Special care has to be taken when scores for these problems are aggregated, as effects may cancel themselves out and appear unproblematic even though the individual challenge scores would be high. 

The questions regarding the degree of virtualization only allowed for binary Yes/No answers. Different degrees and nuances of virtualization aspects could not be expressed, e.g. the degrees of an offshore relation can be investigated using more fine-grained dimensions as described in the taxonomy for GSE by Šmite et al \cite{vsmite2014empirically}.

The survey did not ask for religion, organizational culture, processes used in the company and so on. These aspects that play a large role in the context of a work situation may influence challenges strongly and contribute to a lack of identified links i.e. for the human values. In future research, we plan to incorporate agile vs plan-driven processes and identify whether and how they contribute to challenge occurrences.

We did run a pilot survey to reduce the risk of the questions being ambiguous, however, the anonymous nature of the survey prevented us from reaching out to participants to validate the answers. 

Finally, we realised the notion of Criticality was not explained sufficiently well, and not separated clearly enough from Frequency. Both scores were correlated with a correlation of 0.48 for the team challenges. Respondents were unsure whether they should report on Criticality as it happened in their circumstances, or judge "how critical it was if it happened frequently". For future research, we will focus solely on Frequency plus the concrete implications of a challenge that is not overcome, and abandon the concept of a "general badness".

\section{Conclusion}
We have introduced a compact, refined set of human challenges tailored to software engineering teams. This set addresses the questions of what human challenges occur in SE teams (RQ1). We found that individuals' values do not largely influence the frequency and criticality of challenges within a team (RQ2). However, diversity in nationalities (RQ3) and a higher degree of team virtualization (RQ4) do seem to have an influence in the form of more frequent and critical challenges.

We see various ways of employing the set of human challenges. Additional links can be examined for whether they have an impact on the occurrence of challenges: Agile versus traditional ways of organizing teams, using different means of communication or the existence of special roles such as mentors and mediators in the company.
Specific tools could be tested for their impact on specific challenges or their categories: Tools that allow feedback for employees, easy collaboration or documentation.
The proposed set can also be used to probe the condition of teams and compare them with each other. If teams report a lot of challenges or differ greatly in their reports, mitigation strategies should be considered or could be shared across teams. In this survey, we have collected mitigation strategies for each challenge of the initial set from the respondents. They can be found in appendix \ref{appendix_miti}.

While we found evidence that having multiple nations in a team makes some challenges less frequent, we did not find any significant links between human values and challenges. Future research should investigate which implications of different nationalities produce these results. An in-depth analysis could look at whether there are specific nationalities that work well together. What do they share, what sets them apart? It could also be that companies that have a diverse workforce are more open and tolerant to begin with, contributing to a better work atmosphere, in turn resulting in less frequent challenges.

Finally, in our data, degree of virtualization including digital communication seemed to have an impact on many of the challenges. In our future research, we will examine this on a deeper level: Are there any means of communication or specific tools that have a positive impact on our list of challenges? Is there a difference between text-based communication and being able to see the coworkers' faces? The following constitute more detailed hypotheses that we plan to investigate in our future research:

\begin{itemize}
    \item \textbf{H1.} Video and audio-only both lead to more interpersonal challenges compared to in-person meetings.
    \item \textbf{H2.} Text-based communication leads to more interpersonal challenges than audio communication.
    \item \textbf{H3.} Audio-based communication leads to more interpersonal challenges than communication with both video and audio communication.
    \item \textbf{H4.} The challenge increase stemming from offshore team setups, working remotely and being geographically dispersed can be explained by the increased use of digital communication.
    \item \textbf{H5.} Recomposing teams often leads to more inter-personal challenges.
    \item \textbf{H6.} Having more than one nationality in the team leads to less interpersonal challenges.
    \item \textbf{H7.} Having more interpersonal challenges lead to more intrapersonal challenges.
\end{itemize}

H1-H5 are based on our findings described in section \ref{subTeamVirt}. For H1-H3, we speculate that some problems occur as a result of taking away non-textual cues such as intonation or facial expressions \cite{sherman2013effects}. H6 is a result of our findings of section \ref{subNatDiv}. Finally, we suspect that the presence of many challenges in the working environment or between colleagues have a negative impact and thus lead to an increase of intrapersonal challenges due to stress contagion effects \cite{wethington2000contagion}.

Software engineering is a human activity. Faced with crises like climate change and pandemics, affecting both the way we live and the way we work, research on human challenges in the context of new working styles seems more relevant than ever.
\ifCLASSOPTIONcaptionsoff
  \newpage
\fi


\bibliographystyle{IEEEtran}
\bibliography{biblio}

\begin{thebibliography}{10}
\providecommand{\url}[1]{#1}
\csname url@samestyle\endcsname
\providecommand{\newblock}{\relax}
\providecommand{\bibinfo}[2]{#2}
\providecommand{\BIBentrySTDinterwordspacing}{\spaceskip=0pt\relax}
\providecommand{\BIBentryALTinterwordstretchfactor}{4}
\providecommand{\BIBentryALTinterwordspacing}{\spaceskip=\fontdimen2\font plus
\BIBentryALTinterwordstretchfactor\fontdimen3\font minus
  \fontdimen4\font\relax}
\providecommand{\BIBforeignlanguage}[2]{{%
\expandafter\ifx\csname l@#1\endcsname\relax
\typeout{** WARNING: IEEEtran.bst: No hyphenation pattern has been}%
\typeout{** loaded for the language `#1'. Using the pattern for}%
\typeout{** the default language instead.}%
\else
\language=\csname l@#1\endcsname
\fi
#2}}
\providecommand{\BIBdecl}{\relax}
\BIBdecl

\bibitem{graziotin2014happy}
D.~Graziotin, X.~Wang, and P.~Abrahamsson, ``Happy software developers solve
  problems better: psychological measurements in empirical software
  engineering,'' \emph{PeerJ}, vol.~2, p. e289, 2014.

\bibitem{capretz2014bringing}
L.~F. Capretz, ``Bringing the human factor to software engineering,''
  \emph{IEEE software}, vol.~31, no.~2, pp. 104--104, 2014.

\bibitem{Whittle2019}
J.~Whittle, M.~A. Ferrario, W.~Simm, and W.~Hussain, ``{A Case for Human Values
  in Software Engineering},'' \emph{IEEE Software}, vol. 7459, pp. 1--11, 2019.

\bibitem{fazli2017cultural}
F.~Fazli and E.~A.~C. Bittner, ``Cultural influences on collaborative work in
  software engineering teams,'' in \emph{Proceedings of the 50th Hawaii
  international conference on system sciences}, 2017.

\bibitem{Lenberg2015}
P.~Lenberg, R.~Feldt, and L.~G. Wallgren, ``{Human factors related challenges
  in software engineering - An industrial perspective},'' \emph{Proceedings -
  8th International Workshop on Cooperative and Human Aspects of Software
  Engineering, CHASE 2015}, pp. 43--49, 2015.

\bibitem{Menezes2018}
L.~Menezes and R.~Prikladnicki, ``{Diversity in software engineering},''
  \emph{Proceedings - International Conference on Software Engineering}, pp.
  45--48, 2018.

\bibitem{berg2012anonymity}
R.~W. Berg, ``The anonymity factor in making multicultural teams work: Virtual
  and real teams,'' \emph{Business Communication Quarterly}, vol.~75, no.~4,
  pp. 404--424, 2012.

\bibitem{Conboy2011}
K.~Conboy, S.~Coyle, X.~Wang, and M.~Pikkarainen, ``{People over process: Key
  challenges in agile development},'' \emph{IEEE Software}, vol.~28, no.~4, pp.
  48--57, 2011.

\bibitem{wagner2018systematic}
S.~Wagner and M.~Ruhe, ``A systematic review of productivity factors in
  software development,'' \emph{CoRR}, vol. abs/1801.06475, 2018.

\bibitem{mendez2016naming}
D.~M{\'e}ndez~Fern{\'a}ndez, S.~Wagner, M.~Kalinowski, M.~Felderer, P.~Mafra,
  A.~Vetr{\`o}, T.~Conte, M.-T. Christiansson, D.~Greer, C.~Lassenius
  \emph{et~al.}, ``Naming the pain in requirements engineering: contemporary
  problems, causes, and effects in practice,'' \emph{Empirical Software
  Engineering}, 2016.

\bibitem{han2016framing}
S.~J. Han and M.~Beyerlein, ``Framing the effects of multinational cultural
  diversity on virtual team processes,'' \emph{Small group research}, vol.~47,
  no.~4, pp. 351--383, 2016.

\bibitem{de2015open}
D.~De~Paoli and A.~Ropo, ``Open plan offices--the response to leadership
  challenges of virtual project work?'' \emph{Journal of Corporate Real
  Estate}, 2015.

\bibitem{schwartz2012refining}
S.~H. Schwartz, J.~Cieciuch, M.~Vecchione, E.~Davidov, R.~Fischer,
  C.~Beierlein, A.~Ramos, M.~Verkasalo, J.-E. L{\"o}nnqvist, K.~Demirutku
  \emph{et~al.}, ``Refining the theory of basic individual values.''
  \emph{Journal of personality and social psychology}, vol. 103, no.~4, p. 663,
  2012.

\bibitem{storey2020software}
M.-A. Storey, N.~A. Ernst, C.~Williams, and E.~Kalliamvakou, ``The who, what,
  how of software engineering research: a socio-technical framework,''
  \emph{Empirical Software Engineering}, vol.~25, no.~5, pp. 4097--4129, 2020.

\bibitem{beecham2013we}
S.~Beecham, P.~O’Leary, S.~Baker, I.~Richardson, and J.~Noll, ``Who are we
  doing global software development research for?'' in \emph{8th IEEE
  international conference on global software engineering (ICGSE’13), Bari,
  Italy}, 2013.

\bibitem{yugendhar2017comprehensive}
A.~Yugendhar and P.~Kumar, ``A comprehensive analysis on cultural diversity,''
  \emph{International Journal of Engineering and Management Research (IJEMR)},
  vol.~7, no.~5, pp. 194--201, 2017.

\bibitem{foster2018students}
D.~Foster, F.~Gilardi, P.~Martin, W.~Song, D.~Towey, and A.~White, ``Students
  as co-producers in a multidisciplinary software engineering project:
  addressing cultural distance and cross-cohort handover,'' \emph{Teachers and
  Teaching}, vol.~24, no.~7, pp. 840--853, 2018.

\bibitem{wang2017cultural}
Y.~Wang, J.~Markkula, and J.~Jiang, ``Cultural factors influencing
  international collaborative software engineering education in china,'' in
  \emph{2017 24th Asia-Pacific Software Engineering Conference (APSEC)}.\hskip
  1em plus 0.5em minus 0.4em\relax IEEE, 2017, pp. 31--40.

\bibitem{huang2007cultural}
H.~Huang and E.~M. Trauth, ``Cultural influences and globally distributed
  information systems development: experiences from chinese it professionals,''
  in \emph{Proceedings of the 2007 ACM SIGMIS CPR conference on Computer
  personnel research: The global information technology workforce}.\hskip 1em
  plus 0.5em minus 0.4em\relax ACM, 2007, pp. 36--45.

\bibitem{casey2004practical}
V.~Casey and I.~Richardson, ``Practical experience of virtual team software
  development,'' \emph{European Software Process Improvement}, 2004.

\bibitem{casey2009leveraging}
V.~Casey, ``Leveraging or exploiting cultural difference?'' in \emph{2009
  Fourth IEEE International Conference on Global Software Engineering}.\hskip
  1em plus 0.5em minus 0.4em\relax IEEE, 2009, pp. 8--17.

\bibitem{borchers2003software}
G.~Borchers, ``The software engineering impacts of cultural factors on
  multi-cultural software development teams,'' in \emph{Proceedings of the 25th
  international conference on Software engineering}.\hskip 1em plus 0.5em minus
  0.4em\relax IEEE Computer Society, 2003, pp. 540--545.

\bibitem{alsanoosy2018cultural}
T.~Alsanoosy, M.~Spichkova, and J.~Harland, ``Cultural influences on
  requirements engineering process in the context of saudi arabia,'' in
  \emph{Proceedings of the 13th International Conference on Evaluation of Novel
  Approaches to Software Engineering - Volume 1: ENASE}, INSTICC.\hskip 1em
  plus 0.5em minus 0.4em\relax SciTePress, 2018, pp. 159--168.

\bibitem{tenzer2015leading}
H.~Tenzer and M.~Pudelko, ``Leading across language barriers: Managing
  language-induced emotions in multinational teams,'' \emph{The Leadership
  Quarterly}, vol.~26, no.~4, pp. 606--625, 2015.

\bibitem{tenzer2017influence}
------, ``The influence of language differences on power dynamics in
  multinational teams,'' \emph{Journal of World Business}, vol.~52, no.~1, pp.
  45--61, 2017.

\bibitem{neeley2015global}
T.~Neeley, ``Global teams that work,'' \emph{Harvard Business Review}, vol.~93,
  no.~10, pp. 74--81, 2015.

\bibitem{paasivaara2010practical}
M.~Paasivaara, N.~H. af~Orn{\"a}s, P.~Hynninen, C.~Lassenius, T.~Niinim{\"a}ki,
  and A.~Piri, ``Practical guide to managing distributed software development
  projects,'' \emph{Aalto University, School of Science and Technology, Espoo},
  2010.

\bibitem{liang2007effect}
T.-P. Liang, C.-C. Liu, T.-M. Lin, and B.~Lin, ``Effect of team diversity on
  software project performance,'' \emph{Industrial Management \& Data Systems},
  vol. 107, no.~5, pp. 636--653, 2007.

\bibitem{henderson2018cultural}
L.~S. Henderson, R.~W. Stackman, and R.~Lindekilde, ``Why cultural intelligence
  matters on global project teams,'' \emph{International Journal of Project
  Management}, vol.~36, no.~7, pp. 954--967, 2018.

\bibitem{klitmoller2015speaking}
A.~Klitm{\o}ller, S.~C. Schneider, and K.~Jonsen, ``Speaking of global virtual
  teams: language differences, social categorization and media choice,''
  \emph{Personnel Review}, 2015.

\bibitem{almomani2018empirical}
M.~A. Almomani, S.~Basri, and A.~R. Gilal, ``Empirical study of software
  process improvement in malaysian small and medium enterprises: The human
  aspects,'' \emph{Journal of Software: Evolution and Process}, vol.~30,
  no.~10, p. e1953, 2018.

\bibitem{storey2019towards}
M.-A. Storey, T.~Zimmermann, C.~Bird, J.~Czerwonka, B.~Murphy, and
  E.~Kalliamvakou, ``Towards a theory of software developer job satisfaction
  and perceived productivity,'' \emph{IEEE Transactions on Software
  Engineering}, 2019.

\bibitem{spichkova2015human}
M.~Spichkova, H.~Liu, M.~Laali, and H.~W. Schmidt, ``Human factors in software
  reliability engineering,'' \emph{arXiv preprint arXiv:1503.03584}, 2015.

\bibitem{house2004culture}
R.~J. House, P.~J. Hanges, M.~Javidan, P.~W. Dorfman, and V.~Gupta,
  \emph{Culture, leadership, and organizations: The GLOBE study of 62
  societies}.\hskip 1em plus 0.5em minus 0.4em\relax Sage publications, 2004.

\bibitem{lewis1999cross}
R.~D. Lewis, \emph{Cross cultural communication: A visual approach}.\hskip 1em
  plus 0.5em minus 0.4em\relax Transcreen Publications, 1999.

\bibitem{hofstede2005cultures}
G.~Hofstede, G.~J. Hofstede, and M.~Minkov, \emph{Cultures and organizations:
  Software of the mind}.\hskip 1em plus 0.5em minus 0.4em\relax Citeseer, 2005,
  vol.~2.

\bibitem{brewer2012misuse}
P.~Brewer and S.~Venaik, ``On the misuse of national culture dimensions,''
  \emph{International Marketing Review}, vol.~29, no.~6, pp. 673--683, 2012.

\bibitem{mcsweeney2013fashion}
B.~McSweeney, ``Fashion founded on a flaw: The ecological mono-deterministic
  fallacy of hofstede, globe, and followers,'' \emph{International Marketing
  Review}, vol.~30, no.~5, pp. 483--504, 2013.

\bibitem{Bradner2003}
E.~Bradner, G.~Mark, and T.~D. Hertel, ``{Effects of team size on
  participation, awareness, and technology choice in geographically distributed
  teams},'' \emph{Proceedings of the 36th Annual Hawaii International
  Conference on System Sciences, HICSS 2003}, 2003.

\bibitem{Aube2011}
C.~Aub{\'{e}}, V.~Rousseau, and S.~Tremblay, ``{Team Size and Quality of Group
  Experience: The More the Merrier?}'' \emph{Group Dynamics}, vol.~15, no.~4,
  pp. 357--375, 2011.

\bibitem{Luckhardt2019}
A.~Luckhardt, ``{Analysis of Human Errors in Requirements Engineering},''
  Technische Universit{\"{a}}t M{\"{u}}nchen, Tech. Rep., 2019.

\bibitem{Flint2016}
\BIBentryALTinterwordspacing
M.~Flint, ``{10 Common Problems Project Teams Face},'' 2016. [Online].
  Available:
  \url{https://www.apm.org.uk/blog/10-common-problems-project-teams-face/}
\BIBentrySTDinterwordspacing

\bibitem{Renee2019}
\BIBentryALTinterwordspacing
M.~Renee, ``{Top 10 Diversity Issues at Work},'' 2019. [Online]. Available:
  \url{https://smallbusiness.chron.com/top-10-diversity-issues-work-24939.html}
\BIBentrySTDinterwordspacing

\bibitem{Hammond2016}
\BIBentryALTinterwordspacing
K.~Hammond, ``{Your Slow Decisions Are Frustrating Your Team},'' 2016.
  [Online]. Available:
  \url{https://thinkshiftinc.com/blog/empowering-your-team-to-make-better-marketing-decisions}
\BIBentrySTDinterwordspacing

\bibitem{Batista2019}
\BIBentryALTinterwordspacing
E.~Batista, ``{Work Style Differences},'' 2019. [Online]. Available:
  \url{https://www.edbatista.com/2019/05/work-style-differences.html}
\BIBentrySTDinterwordspacing

\bibitem{holm1979}
S.~Holm, ``A simple sequentially rejective multiple test procedure,''
  \emph{Scandinavian journal of statistics}, pp. 65--70, 1979.

\bibitem{SHAPIRO1965}
\BIBentryALTinterwordspacing
S.~S. Shapiro and M.~B. Wilk, ``An analysis of variance test for normality,''
  \emph{Biometrika}, vol.~52, no. 3-4, pp. 591--611, dec 1965. [Online].
  Available: \url{https://doi.org/10.1093/biomet/52.3-4.591}
\BIBentrySTDinterwordspacing

\bibitem{bortz2010tests}
J.~Bortz and C.~Schuster, ``Tests zur {\"u}berpr{\"u}fung von
  unterschiedshypothesen,'' in \emph{Statistik f{\"u}r Human-und
  Sozialwissenschaftler}.\hskip 1em plus 0.5em minus 0.4em\relax Springer,
  2010, pp. 117--136.

\bibitem{mann1947test}
H.~B. Mann and D.~R. Whitney, ``On a test of whether one of two random
  variables is stochastically larger than the other,'' \emph{The annals of
  mathematical statistics}, pp. 50--60, 1947.

\bibitem{leffingwell2010agile}
D.~Leffingwell, \emph{Agile software requirements: lean requirements practices
  for teams, programs, and the enterprise}.\hskip 1em plus 0.5em minus
  0.4em\relax Addison-Wesley Professional, 2010.

\bibitem{pudelko2015recent}
M.~Pudelko, B.~S. Reiche, and C.~Carr, ``Recent developments and emerging
  challenges in international human resource management,'' 2015.

\bibitem{schaufeli2018work}
W.~B. Schaufeli, ``Work engagement in europe,'' \emph{Organ Dyn}, vol.~47,
  no.~2, pp. 99--106, 2018.

\bibitem{vsmite2014empirically}
D.~{\v{S}}mite, C.~Wohlin, Z.~Galvi{\c{n}}a, and R.~Prikladnicki, ``An
  empirically based terminology and taxonomy for global software engineering,''
  \emph{Empirical Software Engineering}, vol.~19, no.~1, pp. 105--153, 2014.

\bibitem{sherman2013effects}
L.~E. Sherman, M.~Michikyan, and P.~M. Greenfield, ``The effects of text,
  audio, video, and in-person communication on bonding between friends,''
  \emph{Cyberpsychology: Journal of psychosocial research on cyberspace},
  vol.~7, no.~2, 2013.

\bibitem{wethington2000contagion}
E.~Wethington, ``Contagion of stress,'' in \emph{Advances in group
  processes}.\hskip 1em plus 0.5em minus 0.4em\relax Emerald Group Publishing
  Limited, 2000.

\bibitem{brand2015beyond}
A.~Brand, L.~Allen, M.~Altman, M.~Hlava, and J.~Scott, ``Beyond authorship:
  attribution, contribution, collaboration, and credit,'' \emph{Learned
  Publishing}, vol.~28, no.~2, pp. 151--155, 2015.

\end{thebibliography}
\clearpage
\newpage
\small
\appendices
\newcommand*\rot{\rotatebox{90}}
\section{Contributions}
Table \ref{tab:authorshipdetails} elaborates on the details of authorship in regard to this manuscript. The roles and following descriptions have been taken verbatim from Mendez et al. in \cite{mendez2016naming}, which is an extension of roles described by Brand et al. in \cite{brand2015beyond}.

\begin{itemize}
    \item Conceptualisation: Ideas; formulation or evolution of overarching research goals and aims.
    \item Project Administration: Management and coordination responsibility for the research activity planning and execution.
    \item Methodology: Development or design of methodology; creation of models.
    \item Instrument Design: Development / re-design of the instrument used in this replication.
    \item Data Collection: Data collection as national representative in the respective country using the provided infrastructure.
    \item Data Analysis: Application of textual analysis techniques to study and interpret the data.
    \item Data Curation: Management activities to annotate (produce metadata), scrub data and maintain research data (including software code, where it is necessary for interpreting the data itself) for initial use and later reuse.
    \item Data Visualisation: Preparation, creation and/or presentation of the published work, specifically visualisation/ data presentation.
    \item Writing - Original Draft: Preparation, creation and/or presentation of the published work, specifically writing the initial draft (including substantive translation).
    \item Data - Review \& Editing: Preparation, creation and/or presentation of the published work by those from the original research group, specifically critical review, commentary or revision including pre- or post-publication stages.
\end{itemize}
\begin{center}
\caption{Authorship details: Contributions.}
\label{tab:authorshipdetails}
\begin{xtabular}{p{0.45\linewidth}cccc}
\midrule
Role & \rot{M. Hoffmann} & \rot{D. Mendez} & \rot{F. Fagerholm} & \rot{A. Luckhardt} \\
\midrule
Conceptualisation               & X & X &   &   \\
Project Administration          & X & X &   &   \\
Methodology                     & X & X & X &   \\
Instrument Design               & X & X & X &   \\
Data Collection                 & X & X &   &   \\
Data Analysis                   & X &   &   & X \\
Data Curation                   & X &   &   &   \\
Data Visualisation              & X &   &   &   \\
Writing - Draft                 & X & X &   &   \\
Writing - Review \& Editing     & X & X & X &   \\
\end{xtabular}
\end{center}

\section{Reported mitigation measures}\label{appendix_miti}
In following is a summary of the reported measures that respondents had in place for each team challenge.
\newline
\newline
T1, \textbf{Lack of experience}:
\begin{itemize}[noitemsep,topsep=0pt]
    \item Trainings and Coachings
    \item Peer Programming, Code Reviews
    \item Mentors
    \item “Onboarding Checklists”
    \item Regular Meetings for Updates and Knowledge Exchange
    \item Installment of Communities of Practices
\end{itemize}
T2, \textbf{Lack of qualification}:
\begin{itemize}[noitemsep,topsep=0pt]
\item Training Offers (Funding for books, online\&offline courses, trainings, certificates)
\item Mentors
\item Possibility to switch team in-house
\item Open Discussions
\item Strict monitoring of conventions being followed
\item Adapting difficulty of tasks to qualification
\end{itemize}
T3, \textbf{Lack of teamwork skills}:
\begin{itemize}[noitemsep,topsep=0pt]
\item Team Building \& Events
\item Pair Programming
\item Retrospectives \& Team Outings
\item Coaches \& Trainings
\item SCRUM
\item Employ software to measure office mood
\item Flat Hierarchies
\end{itemize}
T4, \textbf{Lack of leadership}:
\begin{itemize}[noitemsep,topsep=0pt]
\item Coachings \& Trainings
\item Foster\&Grow Leaders in-house with guidance by seniors
\item Feedback Sessions \& Retrospectives
\item Employ software to measure office mood
\end{itemize}
T5, \textbf{Communication plan is neglected}:
\begin{itemize}[noitemsep,topsep=0pt]
\item Project Management Software
\item Retrospective \& Transparency Talks
\item Commitments to stick to a plan
\item Trainings
\item Employ a defined process
\item Leadership
\item Ensure everyone has access to updates
\item Trainings
\item Schedule regular meetings
\end{itemize}
T6, \textbf{Business needs are neglected}:
\begin{itemize}[noitemsep,topsep=0pt]
\item Establish Key Performance Indicators
\item Regular Updates
\item Strong engagement of stakeholders
\item Backlog Refinement
\item Involve a project manager
\item Ensure staff understands what the holistic direction\&goal of the company is
\item Open Discussions
\item Feedback
\item Regular presentations about positive consequences of engagement and negative consequences of neglect 
\end{itemize}
T7, \textbf{Demotivation}
\begin{itemize}[noitemsep,topsep=0pt]
\item Optimize distribution of tasks
\item Opportunities to relax (Kicker Table, Ping Pong Tables, Gaming Consoles, special free time)
\item Social Support (Direct valuing of the staff, One-on-one talks, open discussions, team events, reachable leads)
\item Visualizing the goals
\item Gifts (Wearables with team logo)
\item Success Stories
\item Leadership
\item Financial incentives
\item Retrospectives
\item Shift focus regularly (change between implementation, research, analysis, technical maintenance)
\item Employ software to measure office mood
\end{itemize}
T8, \textbf{Gold plating}:
\begin{itemize}[noitemsep,topsep=0pt]
\item Establish Key Performance Indicators
\item Backlog Grooming
\item Roadmaps
\item Detailed task specification and clear requirements
\item SCRUM
\item Strict Product Owners
\end{itemize}
T9, \textbf{Missing willingness to adapt to new software}:
\begin{itemize}[noitemsep,topsep=0pt]
\item Democratic decisions
\item Enforcement by management
\item Demo of new software and showcase of the benefits
\item Trainings
\end{itemize}
T10, \textbf{Missing willingness to adapt to new infrastructure}:
\begin{itemize}[noitemsep,topsep=0pt]
\item Enforcement
\item Training
\item Demo of new infrastructure and showcase the benefits
\item Tutorials
\item Democratic decisions
\item Designate enthusiastic team colleague as a go-to person in case of problems
\end{itemize}
T11, \textbf{Missing willingness to adapt to new processes}:
\begin{itemize}[noitemsep,topsep=0pt]
\item Enforcement
\item Training
\item Leadership
\item Analytic decisions based on pros and cons
\item Open Discussions
\end{itemize}
T12, \textbf{Insufficient collaboration}:
\begin{itemize}[noitemsep,topsep=0pt]
\item Team Events
\item Regular Meetings
\item Coachings
\item Peer Programming
\item Retrospectives
\item Leadership
\item Friendly escalation
\item “Technical Council” – Assembly of all teams with representatives to increase understanding of each other
\end{itemize}
T13, \textbf{Insufficient analysis at the beginning of a task}:
\begin{itemize}[noitemsep,topsep=0pt]
\item Team Reviews
\item Commitment to better planning
\item Introduction of formal specification processes
\item SCRUM
\item Retrospectives
\item Intense communication with stakeholders from the start
\item Pre-Grooming Meetings where team members can ask about areas they do not know
\end{itemize}
T14, \textbf{Subjective interpretations of tasks}:
\begin{itemize}[noitemsep,topsep=0pt]
\item Implement a formalized process of defining tasks
\item Project Management Software
\item Team Reviews
\item Regular planning meetings
\item Definitions of Ready
\item SCRUM
\item Open Discussions
\item Delegate task definition to highly experienced colleagues
\item Workshops
\end{itemize}
T15, \textbf{Work is not solution-oriented}:
\begin{itemize}[noitemsep,topsep=0pt]
\item Introduce Key Performance Indicators
\item Status Updates
\item Team Reviews
\item Milestones
\item Definition of precise Acceptance Criteria
\item Leadership
\item Trainings
\end{itemize}
T16, \textbf{Language barriers}:
\begin{itemize}[noitemsep,topsep=0pt]
\item Language Trainings
\item Funding for voluntary language courses
\item All-English Code-Base and Documentation
\item Team Communication Meetings for Language Practise
\item Usage of appropriate communication channels (written vs verbal)
\item Speak slowly and avoid complicated words
\end{itemize}
T17, \textbf{Conflicts of interests at management level}:
\begin{itemize}[noitemsep,topsep=0pt]
\item Discussions
\item Escalation to board of directors
\item Clear and visible definitions of goals per team
\end{itemize}
T18, \textbf{Missing documentation of the project}:
\begin{itemize}[noitemsep,topsep=0pt]
\item Employ a formalized process in which documentation is an integral part
\item “Definition of Done”
\item Frequent reminders
\end{itemize}
T19, \textbf{Information is not made known to the team}:
\begin{itemize}[noitemsep,topsep=0pt]
\item Regular meetings
\item Coaching
\item SCRUM
\item All project documents are public to the involved staff
\item Wikis
\item Leadership
\item Retrospective Meetings
\item Open Chat Channels
\item Encourage transparency
\end{itemize}
T20, \textbf{Misunderstandings in communication}:
\begin{itemize}[noitemsep,topsep=0pt]
\item One-To-One-Meetings for clarification
\item Soft Skill Training
\item Regular meetings
\item Repeating what one understood for the other to confirm
\item Feedback
\item Encouraging honesty and fear elimination
\item Agree on common terms
\item Documenting agreements
\end{itemize}
T21, \textbf{Certain people dominating discussions}:
\begin{itemize}[noitemsep,topsep=0pt]
\item Allotting certain time slots per person in discussions
\item Moderators
\end{itemize}
T22, \textbf{Missing respect in the workplace}:
\begin{itemize}[noitemsep,topsep=0pt]
\item Appreciation Rounds
\item Mediators
\item Moderated discussions
\item Two-Way Feedback Talks with Leadership
\item Team Events
\item Foster Culture of Respect
\item Supervised 1:1 Talks
\end{itemize}
T23, \textbf{Missing acceptance of alternative lifestyles}:
\begin{itemize}[noitemsep,topsep=0pt]
\item Foster open culture
\item Leadership setting good example
\item Flexible work hours
\item Feedback Sessions
\item Team Events
\end{itemize}
T24, \textbf{Harassment}:
\begin{itemize}[noitemsep,topsep=0pt]
\item Private Conversations with Management
\item Actions taken against the aggressor
\item Feedback talks with leaders
\item Offer psychological council
\item Team Building
\end{itemize}
T25, \textbf{Slow decision making}:
\begin{itemize}[noitemsep,topsep=0pt]
\item Frequent meetings
\item Deciding on one responsible person to handle certain decisions
\item Exposing slow answers in ticketing system
\item Encouraging teams to decide on their own given a deadline
\item Friendly Pressure
\end{itemize}
T26, \textbf{Conflicting working styles}:
\begin{itemize}[noitemsep,topsep=0pt]
\item Open Discussions
\item Compromise (e.g. designate special small time frames where people have to be available)
\item Formalized Procedures as a “synchronization point”
\end{itemize}
T27, \textbf{People getting angry in discussion}s:
\begin{itemize}[noitemsep,topsep=0pt]
\item Mediators
\item Stop talks about point of conflict, then do 1:1 talks
\item Encourage problem-oriented and not people-oriented style of discussion
\item Leadership
\item Feedback Sessions
\item Anger Management Trainings
\item Appointed Peace Keepers
\end{itemize}
T28, \textbf{People crying in discussions}:
\begin{itemize}[noitemsep,topsep=0pt]
\item Stop discussion and give everyone time to calm down
\item Avoid Blaming
\item Emotional Management Training
\item Encourage respectful behaviour in any situation
\end{itemize}
T29, \textbf{Frequently changing team members}:
\begin{itemize}[noitemsep,topsep=0pt]
\item Avoid
\item Ensure good onboarding
\item Management aiming for better planning
\item Increase financial incentives
\item Team Building
\end{itemize}
T30, \textbf{People not reporting problems in time}:
\begin{itemize}[noitemsep,topsep=0pt]
\item Daily Meeting
\item Educating colleagues about the impact
\item Foster atmosphere where people are not afraid to mention problems
\item Allowing various communication channels for reports
\item Leadership
\item Culture of Asking
\item Public approach to consequences of problem making: Eliminate fear of losing job, respect,…
\item Feedback Sessions
\item Retrospectives
\end{itemize}
T31, \textbf{Over-Confidence}:
\begin{itemize}[noitemsep,topsep=0pt]
\item Awareness of previous errors
\item Involve multiple persons in a task
\item Leadership
\item Coaching
\item Feedback sessions
\end{itemize}
T32, \textbf{Pressure from client forwarded to SE team}:
\begin{itemize}[noitemsep,topsep=0pt]
\item Introduce “Business Layer” between client and SE team
\item SCRUM
\item Dedicate staff to handle client requests
\item Employ Project Managers
\item Take care to forward positive client feedback
\item Leadership that shields team from client pressure
\item Direct communication from whole team to client, thus increasing trust and involvement
\item Formalized procedures
\item Feedback sessions
\item Prioritisation and Clear Requirements
\end{itemize}
T33, \textbf{Exaggerated seeking of project problems}:
\begin{itemize}[noitemsep,topsep=0pt]
\item Retrospective Meetings
\item Adhere to Product Backlog
\item Focus on Business Value
\item Feedback Sessions
\end{itemize}
\section{Dataset}\label{appendix_repo}
The complete response dataset, list of modifications done to create our refined challenges, as well as descriptions for the refined challenges, can be found here: \newline
\url{https://github.com/NaPiRE/humanSEChallenges2019}
\end{document}